\documentclass[aps,prl,reprint,superscriptaddress,longbibliography,floatfix]{revtex4-2}
\usepackage{amsmath,amssymb,bm,mathtools}
\usepackage{graphicx}
\usepackage{microtype}
\usepackage[none]{hyphenat}
\usepackage{xcolor}
\usepackage{hyperref}
\hypersetup{colorlinks=true,citecolor=blue,urlcolor=blue,linkcolor=blue}
\newcommand{\Tr}{\operatorname{Tr}}
\newcommand{\ee}{\mathrm{e}}

\newcommand{\PottsNmax}{100}
\newcommand{\PottsTcNmax}{0.6193}
\newcommand{\PottsTcReported}{0.626(3)}
\newcommand{\PottsTcFitLow}{0.6258}
\newcommand{\PottsTcFitHigh}{0.6272}
\newcommand{\BlockEllMax}{9}
\newcommand{\BlockShiftLowN}{0.1439}
\newcommand{\BlockShiftHighN}{0.1439}
\newcommand{\BlockShiftFitLow}{0.1067}
\newcommand{\BlockShiftFitHigh}{0.1363}
\newcommand{\BlockQComp}{0.1323}
\newcommand{\BlockQFitLow}{0.0913}
\newcommand{\BlockQFitHigh}{0.1057}
\newcommand{\BlockSurvivalComp}{0.305}
\newcommand{\TEBDDtShift}{0.0020}
\newcommand{\TEBDBondShift}{\ensuremath{5.0\times10^{-11}}}
\newcommand{\IsingCriticalField}{0.497}
\newcommand{\IsingNmax}{400}
\newcommand{\IsingTcNmax}{6.7772}
\newcommand{\IsingTcInf}{6.7980}
\newcommand{\IsingProbabilityFloor}{10^{-12}}

\newcommand{\FigTwoEllMax}{16}
\newcommand{\FigTwoShiftLowN}{0.1327}
\newcommand{\FigTwoShiftHighN}{0.1327}
\newcommand{\FigTwoQComp}{0.1172}
\newcommand{\FigTwoSurvivalComp}{0.153}
\newcommand{\FigTwoSurvivalFirst}{0.483}
\newcommand{\FigTwoShiftFitLow}{0.1097}
\newcommand{\FigTwoShiftFitHigh}{0.1277}
\newcommand{\FigTwoQFitLow}{0.0947}
\newcommand{\FigTwoQFitHigh}{0.1011}

\usepackage{etoolbox,booktabs,placeins}
\hypersetup{hypertexnames=false}
\makeatletter
\let\ArxivOriginalLabel\label
\let\ArxivOriginalRef\ref
\let\ArxivOriginalAuthor\author
\let\ArxivOriginalAffiliation\affiliation
\let\ArxivOriginalEmail\email
\let\ArxivOriginalThanks\thanks
\let\ArxivOriginalAnd\and
\def\ArxivRestorePreamble{}
\def\do#1{%
  \expandafter\let\csname arxiv@pre@\string#1\endcsname#1%
  \edef\ArxivRestorePreamble{\unexpanded\expandafter{\ArxivRestorePreamble}%
    \let\noexpand#1\expandafter\noexpand\csname arxiv@pre@\string#1\endcsname}%
}
\@preamblecmds
\let\do\noexpand
\newcommand\ArxivBeginSupplement{%
  \close@column
  \clearpage
  \onecolumngrid
  \ArxivRestorePreamble
  \let\arxiv@endpreamble\@empty
  \let\arxiv@begindocument\@empty
  \long\def\AtEndPreamble##1{\appto\arxiv@endpreamble{##1}}%
  \long\def\AtBeginDocument##1{\appto\arxiv@begindocument{##1}}%
  \DeclareMicrotypeSet{arxivsupptext}{encoding={OT1,T1,TS1},family={ntxtlf,qhv}}%
  \MT@map@tlist@c\MT@font@sets\MT@fix@font@set
  \UseMicrotypeSet[expansion]{arxivsupptext}%
  \usepackage[T1]{fontenc}%
  \usepackage{newtxtext,newtxmath}%
  \usepackage{xurl}%
  \arxiv@endpreamble
  \arxiv@begindocument
  \@ifundefined{fontaxes@naming@exception}{}{%
    \fontaxes@naming@exception{shape}{{up}{ulc}}{n}}%
  \mathversion{normal}\normalfont\normalsize
  \SetSymbolFont{boldoperators}{normal}{\tx@enc}{\rmdefaultB}{\bold@wt}{n}%
  \SetSymbolFont{boldoperators}{bold}{\tx@enc}{\rmdefaultB}{\bold@wt}{n}%
  \SetSymbolFont{boldletters}{normal}{OML}{ntxmi}{b}{it}%
  \SetSymbolFont{boldletters}{bold}{OML}{ntxmi}{b}{it}%
  \SetSymbolFont{boldsymbols}{normal}{LMS}{ntxsy}{b}{n}%
  \SetSymbolFont{boldsymbols}{bold}{LMS}{ntxsy}{b}{n}%
  \let\ArxivSavedBMGeneral\bm@general
  \bm@setup{bold}\@ne
  \let\bm@general\ArxivSavedBMGeneral
  \mathversion{normal}%
  \hyphenpenalty=50\exhyphenpenalty=50
  \emergencystretch=0pt\clubpenalty=150\widowpenalty=150
  \let\set@footnotewidth\set@footnotewidth@one
  \let\compose@footnotes\compose@footnotes@one
  \ltx@footnote@pop
  \@booleanfalse\twocolumn@sw
  \@booleanfalse\footinbib@sw
  \@booleanfalse\titlepage@sw
  \long\def\title@column##1{\minipagefootnote@init ##1\minipagefootnote@foot}%
  \frontmatter@init
  \global\setbox\absbox\box\voidb@x
  \let\@received\@empty\let\@revised\@empty
  \let\@accepted\@empty\let\@published\@empty
  \let\author\ArxivOriginalAuthor
  \let\affiliation\ArxivOriginalAffiliation
  \let\email\ArxivOriginalEmail
  \let\thanks\ArxivOriginalThanks
  \let\and\ArxivOriginalAnd
  \let\maketitle\frontmatter@maketitle
  \def\label##1{\ifstrequal{##1}{FirstPage}{\ArxivOriginalLabel{supp:FirstPage}}{\ifstrequal{##1}{LastBibItem}{\ArxivOriginalLabel{supp:LastBibItem}}{\ArxivOriginalLabel{##1}}}}%
  \def\ref##1{\ifstrequal{##1}{LastBibItem}{\ArxivOriginalRef{supp:LastBibItem}}{\ArxivOriginalRef{##1}}}%
  \patchcmd{\frontmatter@makefntext}{frontmatter.}{supp.frontmatter.}{}{}%
  \patchcmd{\frontmatter@footnotemark}{frontmatter.}{supp.frontmatter.}{}{}%
  \patchcmd{\present@bibnote}{frontmatter.}{supp.frontmatter.}{}{}%
  \setcounter{equation}{0}\setcounter{figure}{0}\setcounter{table}{0}%
  \setcounter{section}{0}\setcounter{subsection}{0}\setcounter{footnote}{0}%
  \setcounter{page}{1}%
}
\makeatother

\begin{document}

\title{Phase space anatomy of dynamical quantum phase transitions}
\author{Zakaria Mzaouali}
\email{z.mzaouali@extern.fz-juelich.de}
\affiliation{J\"ulich Supercomputing Centre, Forschungszentrum J\"ulich GmbH, 52425 J\"ulich, Germany}
\affiliation{Institut für Theoretische Physik, Eberhard Karls Universität Tübingen, Auf der Morgenstelle 14, 72076 Tübingen, Germany}
\date{\today}

\begin{abstract}
Dynamical quantum phase transitions (DQPT) are commonly defined either by the behavior of a late-time order parameter or by nonanalyticities in a quantum state's return rate. We show that discrete phase space fundamentally separates these two notions by the scale of information they require. An order parameter transition is determined by a local reduced state, requiring no quasiprobability negativity. In contrast, general global returns can contain information absent from every proper reduced state. For stabilizer returns, we derive an exact decomposition of the rate into distinct costs from loss of support and destructive quantum interference. Dynamics of a qutrit Potts chain show that selective cancellation can reverse the ranking of competing block returns and delay their exchange. An exact control also exhibits a thermodynamic return cusp with zero negativity at the crossing. The existence of a return singularity and the role of interference in selecting its branches are therefore distinct physical questions.
\end{abstract}

\maketitle

\emph{Introduction.}
Quantum quenches extend phase transitions beyond equilibrium, but different observables define different notions of dynamical criticality. A transition of type I (DQPT I) concerns a nonanalytic change of a late time order parameter, as explored in early studies of collective dynamics~\cite{Sciolla2010}. A transition of type II (DQPT II) concerns a singular Loschmidt return rate in the thermodynamic limit, connecting quantum evolution to the analytic structure of a partition function~\cite{Heyl2013,Heyl2018}. Experiments with trapped ions have observed signatures of return singularities and transitions in dynamical order~\cite{Jurcevic2017,Zhang2017}. Their physical interpretation requires understanding what each diagnostic reveals about the evolving state.

Connections between dynamical order and return singularities have been explored in long range Ising chains~\cite{Zunkovic2018}, collective models~\cite{Corps2023}, and systems exhibiting quantum many-body scars, where return exchanges within a degenerate manifold can decouple from order parameter zeros~\cite{PRR_2023_Halimeh}. Signatures of these transitions have also been linked to the growth of quasilocal string operators~\cite{PRL_2021_Dutta} and the dynamics of entanglement~\cite{PRL_2021_Maksym}. To understand what actually distinguishes these two transitions, we must identify the specific physical mechanism that drives the global singularities. This motivates a precise question regarding the return measurement: does the winning branch dominate because it carries more unsigned phase space weight, or because it survives destructive quantum interference more effectively?

Discrete Wigner representations express states and observables on the same finite phase space~\cite{Wootters1987}. Reduced Wigner functions can detect equilibrium transitions in spin chains~\cite{Mzaouali2019,PRE_2024_Hawary}. Studies of the Potts chain find enhanced mana, the logarithm of total absolute Wigner weight, at criticality~\cite{White2021}, and numerical evidence for universal logarithmic scaling of mutual mana~\cite{Tarabunga2024}. The Potts chain also exhibits return singularities after a quench~\cite{Karrasch2017}. Together these observations motivate resolving how the nonclassical structure of a state contributes to a particular dynamical return.

The computational perspective makes this question especially concrete. Approximate phase space dynamics captures transitions in late time Ising magnetization~\cite{Khasseh2020}, while Wigner negativity is a resource for stabilizer computation in odd prime dimension~\cite{Veitch2014} and controls cancellation in quasiprobability sampling~\cite{Pashayan2015}. A total negativity measure, however, does not specify which return branch is suppressed or whether that suppression changes the dominant sector.

Here we show that the existence of a return cusp and the role of interference in selecting its branches are distinct physical questions. We use canonical qutrit phase space, as its Clifford covariant representation ensures that stabilizer reference states have nonnegative Wigner functions on affine supports.~\cite{Gross2006,Zhu2016}. Resolving the evolved state's signed and unsigned weights on these supports identifies when cancellation reverses the ranking of competing returns. An exact product-state control develops a thermodynamic cusp with zero negativity at the crossing, whereas interacting Potts blocks exhibit a delay because cancellation suppresses the competitor more strongly. Combined with an obstruction to reconstructing general returns from local marginals, these results distinguish the information needed to reconstruct a return from the interference that selects its dominant sector.

\begin{figure*}[t]
 \centering
 \includegraphics[width=0.98\textwidth]{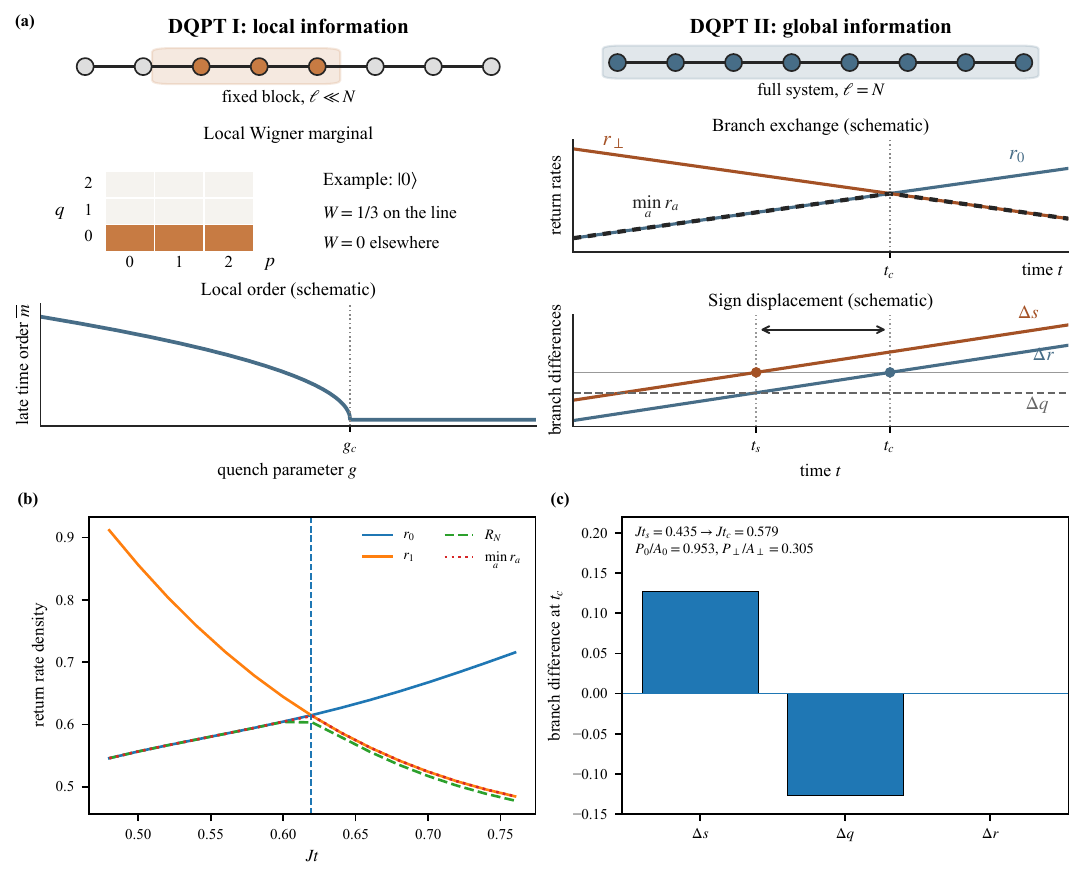}
 \caption{\textbf{Information scale and interference cost.} (a) A fixed block Wigner marginal determines local order, whereas global return branches probe the full system. The qutrit $|0\rangle$ example has positive Wigner weight $1/3$ on one phase space line. The curves are schematic: late time order changes with quench strength $g$, while smooth return branches exchange in time and form a cusp in their lower envelope. Writing each rate as $r_a=s_a+q_a$ separates unsigned support from cancellation; unequal branch costs can shift the exchange from the unsigned crossing $t_s$ to the physical crossing $t_c$. (b,c) Numerical results for the open Potts chain quenched from $|0\rangle^{\otimes N}$ to $h/J=1.5$. (b) Global rates at $N=100$: $r_0$ is the initial branch and $r_1=r_2$ are symmetry related competitors. Their dotted lower envelope has a cusp at $Jt_c=\PottsTcNmax$ (vertical line), while $R_N=-N^{-1}\ln(P_0+P_1+P_2)$ remains smooth. (c) Initial minus competing differences on the central block $\ell=9$ of the same chain, evaluated at its own crossing $Jt_c\simeq0.579$. Positive $\Delta s$ favors the competitor, but negative $\Delta q$ compensates this advantage, giving $\Delta r=0$. Only $30.5\%$ of its unsigned weight survives, compared with $95.3\%$ for the initial branch; this selective cancellation delays the block exchange from $Jt_s\simeq0.435$.}
 \label{fig:framework}
\end{figure*}

\emph{Information scale.}
Let $\rho_\ell(t)$ be a contiguous block of $\ell$ qutrits and $W_{\rho_\ell}(u)$ its Wigner function. Every observable supported on that block obeys
\begin{equation}
 \langle O_\ell\rangle_t=3^\ell\sum_u W_{\rho_\ell(t)}(u)W_{O_\ell}(u).
 \label{eq:localoverlap}
\end{equation}
A local DQPT I order parameter is thus determined by a Wigner marginal of fixed size. Following the same marginal as $\ell$ grows toward $N$ connects local order to global returns within one mathematical hierarchy [Fig.~\ref{fig:framework}(a)]. At its macroscopic endpoint, a reference state $|\phi_a\rangle$ defines the return branch $P_a^{(N)}(t)=|\langle\phi_a|\psi(t)\rangle|^2$. For orthogonal branches, the total rate is $R_N=-N^{-1}\ln\sum_a P_a^{(N)}$.

\begin{figure*}[t]
 \centering
 \includegraphics[width=0.98\textwidth]{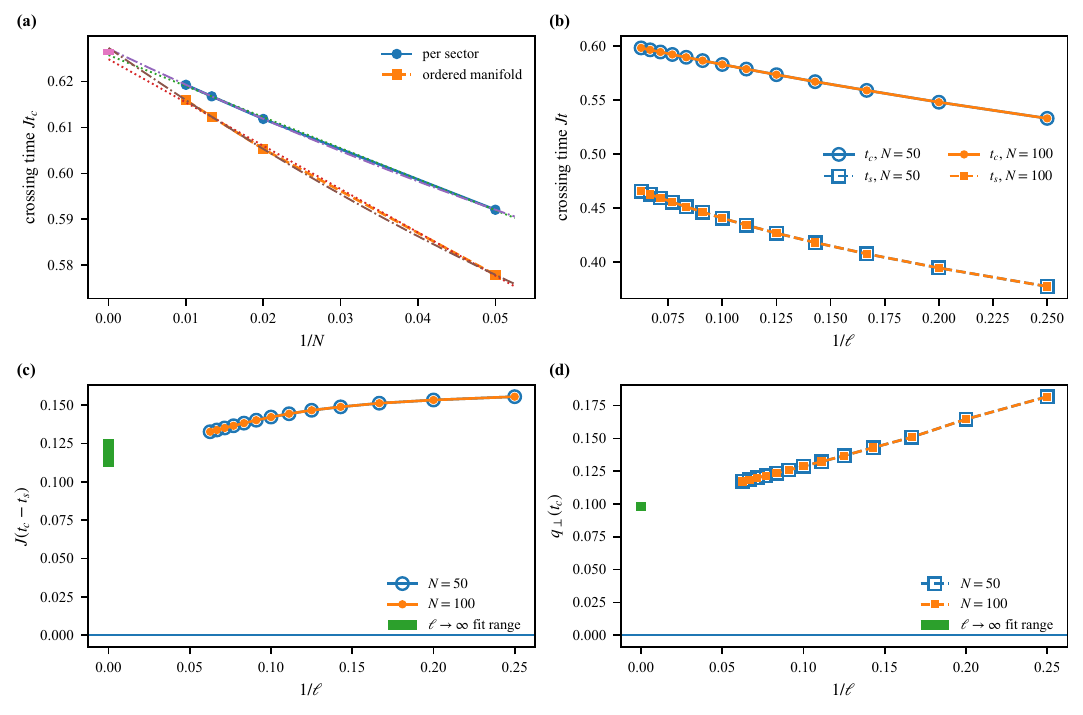}
 \caption{\textbf{Global branch exchange and cancellation on growing blocks.} Open Potts chains are quenched from $|0\rangle^{\otimes N}$ to $h/J=1.5$. (a) Global crossing times versus inverse system size, $N=20,50,75,100$. Circles compare $P_0$ with one competing sector; squares compare it with the grouped weight $P_1+P_2$. Grouping shifts the competing rate by $-(\ln2)/N$, so the finite size crossings differ but approach a common limit. Dotted and dash-dotted curves are linear and quadratic fits; the mark at $1/N=0$ spans the two per-sector intercepts. (b) Physical crossings $t_c$ of the signed returns and auxiliary crossings $t_s$ of their unsigned Wigner weights for central blocks $\ell=4,\ldots,\FigTwoEllMax$, using the grouped competitor $P_\perp=P_1+P_2$ and $A_\perp=A_1+A_2$. Block crossings are located by direct evolution to each trial time, with step $J\delta t=0.02$, cap 32, and cutoff $10^{-9}$. Open blue symbols ($N=50$) and smaller filled orange symbols ($N=100$) overlap, indicating little boundary sensitivity, while $t_c$ and $t_s$ remain separated. (c) The positive delay $J(t_c-t_s)$ shows that cancellation postpones branch exchange at every measured block size. (d) The competing cost $q_\perp(t_c)=\ell^{-1}\ln(A_\perp/P_\perp)$ decreases with block size, although the surviving fraction $P_\perp/A_\perp$ also decreases. In (c,d), green bars at $1/\ell=0$ span 12 extrapolated intercepts: linear and quadratic fits with lower cutoffs $\ell=4,5,6$, for both chain sizes. These positive ranges support persistence within the tested fit family; they quantify fit sensitivity, carry no statistical confidence level, and do not establish the infinite block limit.}
 \label{fig:scaling}
\end{figure*}

The information obstruction for a general global return is explicit. Define $|\mathrm{GHZ}_\varphi\rangle=(|0\rangle^{\otimes N}+\ee^{i\varphi}|1\rangle^{\otimes N})/\sqrt2$. For any nonempty proper subset $B$, let $A$ be its complement and $|a_A\rangle=|a\rangle^{\otimes|A|}$. Orthogonality on $B$ removes the off diagonal terms, giving
\begin{align}
 \Tr_B|\mathrm{GHZ}_\varphi\rangle\langle\mathrm{GHZ}_\varphi|
 &=\tfrac12\bigl(|0_A\rangle\langle0_A|+|1_A\rangle\langle1_A|\bigr),\nonumber\\
 |\langle\mathrm{GHZ}_0|\mathrm{GHZ}_\varphi\rangle|^2
 &=\cos^2(\varphi/2).
 \label{eq:ghz-proof}
\end{align}
All proper reduced states are independent of $\varphi$, while the global return ranges from zero to one. Thus proper Wigner marginals do not determine general Loschmidt returns. The hierarchy also exposes a familiar thermodynamic analogy: cancellation on a return support carries an interference free energy density.

\begin{figure*}[t]
 \centering
 \includegraphics[width=0.98\textwidth]{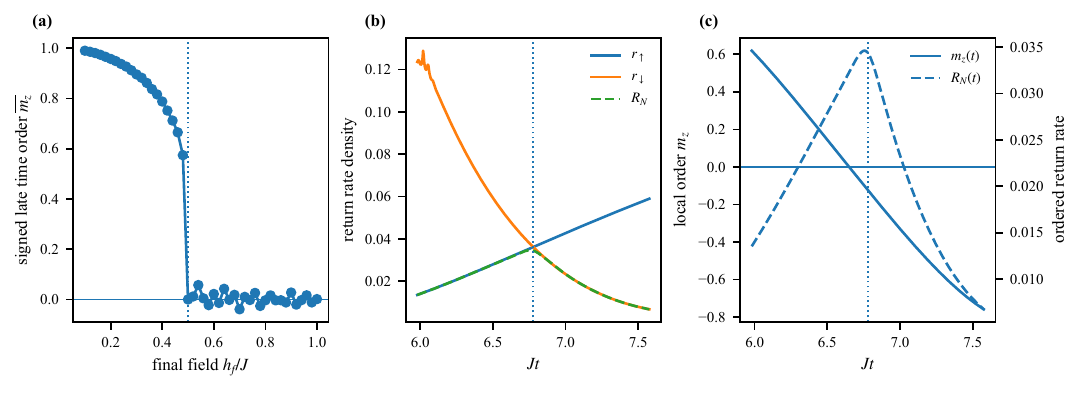}
 \caption{\textbf{Local order and global returns in the collective Ising model.} Dynamics starts fully polarized along $+z$ under $H=-(2J/N)S_z^2-2h_fS_x$. (a) Magnetization averaged over $60\le Jt\le100$ versus final field. The loss of persistent order near the separatrix $h_f/J=1/2$ is the DQPT I diagnostic; the dotted line marks the exact separatrix. At this field the plotted average uses the analytic trajectory $m_z(t)=\operatorname{sech}(Jt)$ to avoid numerical instability; residual oscillations above it reflect the finite averaging window. (b) Exact quantum return rates $r_{\uparrow,\downarrow}=-N^{-1}\ln P_{\uparrow,\downarrow}$ to the two fully polarized states, for $N=400$ and $h_f/J=0.7$. The dotted line marks the exchange at $Jt_c=\IsingTcNmax$. The dashed rate $R_N=-N^{-1}\ln(P_\uparrow+P_\downarrow)$ follows the dominant branch and rounds the crossing at finite $N$. (c) Local magnetization $m_z(t)=2\langle S_z\rangle/N$ (solid, left axis) and $R_N(t)$ (dashed, right axis) for the same quantum evolution as (b). Magnetization passes smoothly through zero at $Jt\simeq6.650$, before the branch exchange, where $m_z\simeq-0.122$. Local symmetry reversal and global return competition are therefore distinct events in this finite system.}
 \label{fig:ising}
\end{figure*}

\emph{Exact interference cost.}
For $|\phi_a^{(\ell)}\rangle=|a\rangle^{\otimes\ell}$, the Wigner function is uniform and positive on an affine support $S_a^{(\ell)}$. The physical return and its unsigned counterpart are
\begin{equation}
 P_a^{(\ell)}=\sum_{u\in S_a^{(\ell)}}W_{\rho_\ell}(u),\qquad
 A_a^{(\ell)}=\sum_{u\in S_a^{(\ell)}}|W_{\rho_\ell}(u)|.
 \label{eq:signedreturn}
\end{equation}
If $\nu_a^{(\ell)}$ denotes the absolute negative weight on that same support, then $A_a^{(\ell)}-P_a^{(\ell)}=2\nu_a^{(\ell)}$. For positive return weight, define
\begin{align}
 r_a^{(\ell)}&=-\ell^{-1}\ln P_a^{(\ell)},\qquad
 s_a^{(\ell)}=-\ell^{-1}\ln A_a^{(\ell)},\nonumber\\
 q_a^{(\ell)}&=\ell^{-1}\ln\bigl(A_a^{(\ell)}/P_a^{(\ell)}\bigr)\ge0.
 \label{eq:decomposition}
\end{align}
These quantities satisfy the exact identity
\begin{equation}
 \boxed{r_a^{(\ell)}=s_a^{(\ell)}+q_a^{(\ell)}},\qquad
 \frac{P_a^{(\ell)}}{A_a^{(\ell)}}=\ee^{-\ell q_a^{(\ell)}}.
 \label{eq:average-sign}
\end{equation}
The ratio is a conditional average sign. Its negative logarithm per site is the interference cost for branch $a$, analogous to the free energy penalty in a sign problem~\cite{Pashayan2015}. The unsigned weight does not need to be a probability; this is an analogy, not an additional equilibrium ensemble. A positive limit of $q_a^{(\ell)}$ means that cancellation suppresses the physical return exponentially with block size. The canonical construction, zero-return conventions, and branch-ordering criteria are detailed in Supplemental Sec.~S\,I~\cite{Supplement}.

For two branches, define initial minus competing differences, so that $\Delta r=\Delta s+\Delta q$. The physical exchange is $\Delta r(t_c)=0$; the auxiliary exchange after removing Wigner signs is $\Delta s(t_s)=0$. Both use the same evolved state and supports. Positivity on both supports around the event makes these differences identical. A displacement $t_c\ne t_s$ between corresponding exchanges therefore witnesses an effect of signs on branch selection. The converse does not hold as equal interference costs cancel in $\Delta q$ even when both are nonzero. Away from ties, opposite rankings occur precisely when $\Delta s\,\Delta r<0$: cancellation then changes which branch has the larger return. This is a criterion for the effect of interference on the specified observable; the unsigned comparison is a diagnostic of the same state, not a second physical evolution.

The branch differences versus time in the lower right of Fig.~\ref{fig:framework}(a) illustrate this displacement. The orange curve is $\Delta s=s_0-s_\perp$, the dashed gray line is $\Delta q=q_0-q_\perp$, and the blue curve is their sum $\Delta r=r_0-r_\perp$. A negative $\Delta r$ favors the initial return, while a positive value favors the competitor. In the illustrated case, $\Delta q<0$ means that stronger cancellation penalizes the competing branch and shifts $\Delta r$ below $\Delta s$. At $t_s$, the unsigned weights are equal, but $\Delta r(t_s)=\Delta q(t_s)<0$, so the initial return still dominates. Throughout $t_s<t<t_c$, unsigned support favors the competitor while the physical return favors the initial branch. The exchange occurs only at $t_c$, where $\Delta s=-\Delta q$; the horizontal arrow marks the resulting delay. The straight curves and constant $\Delta q$ are schematic: actual branch differences can vary nonlinearly with time.

Figure~\ref{fig:framework}(b) illustrates how smooth return branches organize the Potts transition studied below. At $N=100$, the initial branch exchanges dominance with two symmetry related competitors at $Jt_c=\PottsTcNmax$. Their lower envelope has a cusp, while the physical sum retains a smooth rate $R_N$ near the crossing. For three branches, $\min_a r_a-(\ln3)/N\le R_N\le\min_a r_a$. The envelope therefore identifies the branch selection mechanism; establishing a thermodynamic singularity also requires scaling in $N$.

Panel (c) resolves the interference on the central $\ell =9$ site block, whose crossing differs from the global event in (b). At its $Jt_c\simeq0.579$, $\Delta s\simeq0.127$ favors the competing support, but $\Delta q\simeq-0.127$ offsets this advantage, restoring $\Delta r=0$. The initial branch retains $95.3\%$ of its unsigned weight, compared with $30.5\%$ for a competitor. Since $P_0=P_\perp$ at this crossing, the ratio $A_\perp/A_0=\ee^{\ell\Delta s}\simeq3.1$: cancellation compensates an unsigned advantage of more than a factor of three. Removing signs advances the block exchange to $Jt_s\simeq0.435$. At the sampled time $Jt=0.50$ between the crossings, $A_\perp/A_0\simeq1.8$ whereas $P_0/P_\perp\simeq7.2$: the initial block return wins despite the larger unsigned competing weight. Thus interference changes which return dominates, beyond any association between total negativity and the exchange. The following control tests the scope of this mechanism.

\emph{An exact control with zero negativity.}
Negativity is not necessary for every DQPT II. Consider independent qutrits with $H_{\rm rot}=\sum_j h_{{\rm rot},j}$ and
\begin{equation}
 h_{\rm rot}=i\Omega\bigl(|s\rangle\langle0|-|0\rangle\langle s|\bigr),
 \qquad |s\rangle=\frac{|1\rangle+|2\rangle}{\sqrt2}.
 \label{eq:rotation}
\end{equation}
Starting from $|0\rangle^{\otimes N}$, each site evolves to $\cos(\Omega t)|0\rangle+\sin(\Omega t)|s\rangle$. The ordered returns are
\begin{equation}
 P_0=|\cos(\Omega t)|^{2N},\qquad
 P_1=P_2=\left|\frac{\sin(\Omega t)}{\sqrt2}\right|^{2N}.
\end{equation}
At $\Omega t_*=\arccos(1/\sqrt3)$, all rates equal $\ln3$. Their slopes differ: $\partial_t r_0=2\sqrt2\Omega$ and $\partial_t r_{1,2}=-\sqrt2\Omega$. Thus the thermodynamic rate $\lim_{N\to\infty}[-N^{-1}\ln(P_0+P_1+P_2)]=\min_a r_a$ has a cusp. Yet the state at $t_*$ is the stabilizer product $|+\rangle^{\otimes N}$, with $|+\rangle=(|0\rangle+|1\rangle+|2\rangle)/\sqrt3$. Its Wigner function is nonnegative, and $A_a=P_a$, hence $q_a=0$, for every ordered branch. The cusp therefore cannot by itself certify Wigner negativity or entanglement in the critical state: it records a change of the dominant return even in this product evolution. Positivity is asserted at the crossing; the entire trajectory does not need to be positive.

\emph{Interacting Potts chain.}
We now ask whether signs alter an interacting transition. We quench the open Potts chain with three states per site,
\begin{equation}
 H=-J\sum_{j=1}^{N-1}(Z_j^\dagger Z_{j+1}+\mathrm{H.c.})
   -h\sum_{j=1}^{N}(X_j+X_j^\dagger),
 \label{eq:potts}
\end{equation}
from $|0\rangle^{\otimes N}$ to $h/J=1.5$~\cite{Karrasch2017}. Here $Z_j$ and $X_j$ denote the clock and cyclic shift operators on site $j$, defined by $Z|a\rangle=\omega^a|a\rangle$ and $X|a\rangle=|(a+1)\bmod3\rangle$, with $a=0,1,2$ and $\omega=\ee^{2\pi i/3}$. We evolve matrix product states using time evolving block decimation (TEBD) of second order~\cite{Vidal2004,Suzuki1976} and contract all three ordered branches independently. Global scaling reaches $N=\PottsNmax$; central Wigner blocks reach $\ell=\FigTwoEllMax$ in chains of 50 and 100 sites. The evolution scheme and contractions are detailed in Supplemental Sec.~S\,II~\cite{Supplement}.

Fits linear and quadratic in $1/N$ place the crossing at $Jt_c\in[\PottsTcFitLow,\PottsTcFitHigh]$ [Fig.~\ref{fig:scaling}(a)]. The crossing is highly robust against the tested truncation errors: at $N=\PottsNmax$, bond refinement shifts it by less than $\TEBDBondShift$, and Trotter step refinement by $\TEBDDtShift$. Including this step sensitivity gives the conservative estimate $Jt_c=\PottsTcReported$. For symmetry related competitors, grouping $P_1+P_2$ changes the competing rate by exactly $-(\ln2)/N$; the induced time shift also depends on the crossing slope.

The physical and auxiliary block crossings remain distinct [Fig.~\ref{fig:scaling}(b)]. At $\ell=\FigTwoEllMax$, their displacements are $J(t_c-t_s)=\FigTwoShiftLowN$ and $\FigTwoShiftHighN$ for the two chain sizes. Agreement between the two bulk environments supports insensitivity to boundaries over the measured window. The displacement identifies a time interval in which greater unsigned weight fails to predict the dominant return, providing a direct test of the branch selection mechanism illustrated in Fig.~\ref{fig:framework}(a,c).

Figure~\ref{fig:scaling}(c) shows that this delay decreases slowly as the block grows but remains positive. In panel (d), the competing cost also decreases, reaching $q_\perp(t_c)=\FigTwoQComp$. Nevertheless, the surviving fraction falls from $\FigTwoSurvivalFirst$ at $\ell=4$ to $\FigTwoSurvivalComp$ at $\ell=\FigTwoEllMax$: suppression accumulates through the product $\ell q_\perp$ in $P_\perp/A_\perp=\ee^{-\ell q_\perp}$. A decreasing cost per site therefore coexists with stronger cancellation of the return.

The green bars at $1/\ell=0$ summarize the extrapolated intercepts. Each spans 12 fitted intercepts from both chain sizes, two polynomial orders, and three lower block cutoffs. The displacement spans $[\FigTwoShiftFitLow,\FigTwoShiftFitHigh]$ in (c), and the competing cost spans $[\FigTwoQFitLow,\FigTwoQFitHigh]$ in (d). The two scaling tests answer different physical questions: increasing $N$ tests the global return singularity, while increasing $\ell$ tests whether the interference that changes block selection remains relevant on larger supports. A positive limiting $q_\perp$ alone would establish extensive suppression, but its effect on branch ordering depends on $\Delta q$ relative to $\Delta s$. Direct block crossings and their numerical refinements are reported in Supplemental Sec.~S\,III~\cite{Supplement}.

\emph{Ising benchmark.}
The collective Ising Hamiltonian $H=-(2J/N)S_z^2-2hS_x$, starting fully polarized along $+z$, provides a complementary comparison of the two information scales. Permutation symmetry makes $m_z(t)=2\langle S_z\rangle/N$ a one-site expectation value, whereas $P_{\uparrow,\downarrow}$ project onto the two fully polarized many-body states. This setting connects to studies relating dynamical order and return singularities through symmetry and collective dynamics~\cite{Zunkovic2018,Corps2023}.

Figure~\ref{fig:ising}(a) shows the DQPT I diagnostic in classical spin dynamics, corresponding to the thermodynamic limit of the collective model. Below the separatrix $h/J=1/2$, trajectories remain in the initially selected magnetization sector and retain positive signed late time order. Above it, trajectories explore both signs and their long time signed average vanishes. The scan averages over $60\le Jt\le100$; its small residual oscillations above the transition reflect the finite window. At the separatrix, we use the exact trajectory $m_z(t)=\operatorname{sech}(Jt)$, since numerical drift near the unstable saddle can otherwise produce a spurious signed average. This panel varies the quench field and uses classical dynamics, whereas panels (b,c) follow one quantum quench at finite $N$.

In Fig.~\ref{fig:ising}(b), exact evolution in the symmetric spin subspace at $N=\IsingNmax$ and $h/J=0.7$ resolves an exchange at $Jt_c=\IsingTcNmax$. Before this event $r_\uparrow<r_\downarrow$, so the initial ordered return dominates; afterwards the opposite branch dominates. The total rate is smooth and differs from their lower envelope by at most $(\ln2)/N$. The panel thus displays the branch mechanism underlying a DQPT II precursor, rather than an exact finite system nonanalyticity. Both probabilities exceed $\IsingProbabilityFloor$ at both samples bracketing the accepted crossing. The numerical methods and precision criterion are given in Supplemental Sec.~S\,II, and the comparison with local magnetization is detailed in Sec.~S\,IV~\cite{Supplement}.

Figure~\ref{fig:ising}(c) compares local and global observables along that same finite system trajectory. The magnetization crosses zero at $Jt\simeq6.650$ and remains smooth through the later return exchange, where $m_z(t_c)\simeq-0.122$. Meanwhile $R_N$ develops a rounded peak near the exchange. Equal probabilities for the two extreme polarized configurations therefore does not imply equal total weight in the positive and negative magnetization sectors: $m_z$ receives contributions from all intermediate spin projections. This explains why a local zero cannot substitute for the global branch criterion. The finite system offset does not establish a separation of thermodynamic critical points, and a temporal zero of $m_z$ is not itself the field-driven DQPT I of panel (a).

\emph{Discussion.}
The physical conclusion is that a return singularity and its sensitivity to interference are distinct properties of quantum dynamics. The exact control establishes a thermodynamic cusp at a Wigner positive product state. The Potts results establish the complementary possibility: selective cancellation keeps the initial block return dominant after a competing branch has acquired the larger unsigned weight. A cusp alone therefore cannot diagnose negativity at the critical state, while a nonzero displacement between matched signed and unsigned crossings identifies where negativity matters for branch selection. This separates the existence of a dynamical transition from the mechanism that determines its observed return probabilities.

This distinction sharpens the interpretation of competing branch pictures~\cite{PRR_2023_Halimeh} and classifications based on precession and entanglement~\cite{PRL_2021_Maksym}. The additional physical information is which return is penalized by interference and whether that penalty changes the dominant sector. A common cancellation cost can suppress both returns exponentially without changing their ordering; unequal costs can reverse it. Mana, the logarithm of the total absolute Wigner weight over phase space, does not retain this information about the selected measurement~\cite{Veitch2014,White2021,Tarabunga2024}. The branch selection criterion also does not require either cost to be singular: smooth costs can shift the crossing and change the branch slopes. In the lower envelope mechanism, the nonanalyticity arises from the exchange of dominant branches.

The information hierarchy clarifies what local observations can establish. The GHZ example rules out reconstruction of arbitrary global returns from proper marginals; it does not rule out useful local signatures within a specified model. Local return rates and quasilocal strings already reveal such signatures~\cite{Halimeh2021,PRL_2021_Dutta}, and the Ising comparison shows why their relation to global branch exchange requires dynamical information. Here each block probability is itself an expectation of the product of local projectors onto one ordered configuration. Increasing the block length therefore connects accessible joint populations to the global return, while the accompanying interference cost tests whether the same branch is being selected for the same reason. Agreement of a local marker with a return feature does not determine the underlying balance of support and cancellation.

The Potts evidence establishes this change of branch selection on the measured blocks, independently of extrapolation. The agreement between chains of 50 and 100 sites and the stronger accumulated suppression as the block grows make the mechanism relevant to bulk dynamics over the reported window. Establishing its persistence in the thermodynamic return requires larger supports and control of the relative cost $\Delta q$, not only the competing cost $q_\perp$. The positive fitted displacement and cost motivate that test; they do not establish a universal class of DQPTs controlled by negativity. The distinction matters because a global branch exchange can survive even if the interference contribution to its location vanishes with increasing scale.

The same conditional sign connects the physical mechanism to its computational and experimental accessibility. For independent samples on a return support drawn with probability $|W|/A_a$, the sample mean of the sign has relative variance $(\ee^{2\ell q_a}-1)/M$ for $M$ samples~\cite{Troyer2005,Pashayan2015}. Thus a cost that suppresses a selected return also controls the cancellation overhead of this estimator, without implying an algorithm independent hardness bound. Our tensor contractions enumerate the $3^\ell$ cells on each return support without storing all $3^{2\ell}$ Wigner values; the split contraction for the larger blocks is detailed in Supplemental Sec.~S\,II~\cite{Supplement}. Ordered populations and local magnetization already enter experiments on dynamical transitions~\cite{Jurcevic2017,Zhang2017}; estimating local properties with methods such as classical shadows~\cite{Huang2020} suggests a route toward the block hierarchy. Reconstructing the nonlinear unsigned weight and resolving small probabilities remain additional measurement demands. The resulting experimental target is that distinguishing a return branch that wins through greater unsigned weight from one that wins because its competitor suffers stronger cancellation.

\emph{Acknowledgments.}
Z.M. acknowledges funding from the Ministry of Economic Affairs, Labour and Tourism Baden-Württemberg in the frame of the Competence Center Quantum Computing Baden-Württemberg (project ``KQCBW25'').

\emph{Data availability.}
The source code to reproduce the results and all the data used to generate the plots is available in the accompanying GitHub repository \url{https://github.com/MoorishQubit/dqpt-wigner/}

\ArxivBeginSupplement
\graphicspath{{supplement_v08/figures/}}
\renewcommand{\PottsNmax}{100}
\renewcommand{\PottsTcNmax}{0.6193}
\renewcommand{\PottsTcReported}{0.626(3)}
\renewcommand{\PottsTcFitLow}{0.6258}
\renewcommand{\PottsTcFitHigh}{0.6272}
\renewcommand{\BlockEllMax}{9}
\renewcommand{\BlockShiftLowN}{0.1439}
\renewcommand{\BlockShiftHighN}{0.1439}
\renewcommand{\BlockShiftFitLow}{0.1103}
\renewcommand{\BlockShiftFitHigh}{0.1357}
\renewcommand{\BlockQComp}{0.132263}
\renewcommand{\BlockQFitLow}{0.0916}
\renewcommand{\BlockQFitHigh}{0.1055}
\renewcommand{\BlockSurvivalComp}{0.304110}
\renewcommand{\TEBDDtShift}{0.0020}
\renewcommand{\TEBDBondShift}{\ensuremath{5.0\times10^{-11}}}
\renewcommand{\IsingCriticalField}{0.5}
\renewcommand{\IsingNmax}{400}
\renewcommand{\IsingTcNmax}{6.7772}
\renewcommand{\IsingTcInf}{6.7980}
\renewcommand{\IsingProbabilityFloor}{10^{-12}}
\newcommand{\SignedPottsTau}{0.626985981}
\newcommand{\SignedPottsSlope}{1.8984472}
\newcommand{\SignedPottsDt}{0.00125}
\newcommand{\SignedPottsPreviousTau}{0.626991644}
\newcommand{\SignedPottsPreviousDt}{0.0025}
\newcommand{\SignedPottsSlopeSpan}{0.000826054}
\newcommand{\SignedPottsMinModulusGap}{0.743}
\newcommand{\BlockIndividualTau}{0.612792043}
\newcommand{\BlockIndividualDeltaS}{0.048811070}
\newcommand{\BlockIndividualDeltaQ}{-0.048811070}
\newcommand{\BlockIndividualCompetingQ}{0.085670406}
\newcommand{\BlockIndividualSurvival}{0.462534718}
\newcommand{\BlockIndividualUnsignedTau}{0.514588709}
\newcommand{\BlockGroupedTau}{0.578686490}
\newcommand{\BlockGroupedDeltaS}{0.127295281}
\newcommand{\BlockGroupedDeltaQ}{-0.127295281}
\newcommand{\BlockGroupedCompetingQ}{0.132262939}
\newcommand{\BlockGroupedSurvival}{0.304109801}
\newcommand{\BlockGroupedUnsignedTau}{0.434310185}
\newcommand{\BlockFineTau}{0.612442014}
\newcommand{\BlockFineDeltaS}{0.048954110}
\newcommand{\GlobalBlockSixDeltaR}{0.04507236}
\newcommand{\GlobalBlockSixDeltaS}{0.04485949}
\newcommand{\GlobalBlockSixDeltaQ}{0.00021287}
\newcommand{\ConnectedThetaStar}{0.955316618}
\newcommand{\ConnectedGapLower}{0.448970797}
\newcommand{\IsingEarlyNOneHundredTau}{1.402204623}

\renewcommand{\FigTwoEllMax}{16}
\renewcommand{\FigTwoShiftLowN}{0.1327}
\renewcommand{\FigTwoShiftHighN}{0.1327}
\renewcommand{\FigTwoQComp}{0.1172}
\renewcommand{\FigTwoSurvivalComp}{0.153}
\renewcommand{\FigTwoSurvivalFirst}{0.483}
\renewcommand{\FigTwoShiftFitLow}{0.1097}
\renewcommand{\FigTwoShiftFitHigh}{0.1277}
\renewcommand{\FigTwoQFitLow}{0.0947}
\renewcommand{\FigTwoQFitHigh}{0.1011}

\renewcommand{\Tr}{\operatorname{Tr}}
\newcommand{\W}{W}
\newcommand{\supportrate}{s}
\newcommand{\signcost}{q}
\setcounter{secnumdepth}{2}
\renewcommand{\thesection}{S\,\Roman{section}}
\renewcommand{\theequation}{S\arabic{equation}}
\renewcommand{\thefigure}{S\arabic{figure}}
\renewcommand{\thetable}{S\Roman{table}}

\title{Supplemental Material: Phase Space Anatomy of Dynamical Quantum Phase Transitions}
\author{Zakaria Mzaouali}
\email{z.mzaouali@extern.fz-juelich.de}
\affiliation{J\"ulich Supercomputing Centre, Forschungszentrum J\"ulich GmbH, 52425 J\"ulich, Germany}
\affiliation{Institut für Theoretische Physik, Eberhard Karls Universität Tübingen, Auf der Morgenstelle 14, 72076 Tübingen, Germany}
\date{\today}
\maketitle

\section{Canonical phase space and return rates}
\label{sec:scope}\label{sec:qutritphase}\label{sec:branches}\label{sec:sign-decomposition}
This Supplemental Material gives the phase-space identities and numerical
checks supporting the main text. We first define the canonical Wigner
representation, the return-rate decomposition, and the criterion for
cancellation to change branch ordering. Section~\ref{sec:potts-numerics}
describes the Potts and collective Ising numerical methods.
Section~\ref{sec:potts-results} reports the infinite-chain signed crossing
and finite-block comparisons. Section~\ref{sec:ising-benchmark} relates
local magnetization to global return competition in the collective Ising
model.

We use $N$ for global length, $\ell$ for a reduced block, $\tau=Jt$ for Potts
and Ising, and $\theta=\Omega t$ for the exact construction.

\subsection{Canonical representation}

For one qutrit, with all indices in $\mathbb Z_3$, define
\begin{equation}
 X|j\rangle=|j+1\rangle,\quad Z|j\rangle=\omega^j|j\rangle,\quad
 \omega=e^{2\pi i/3},\qquad
 T(q,p)=\omega^{-2^{-1}qp}Z^pX^q,\quad 2^{-1}=2.
 \label{eq:qutrit-weyl}
\end{equation}
We distinguish the phase point operator $\mathsf A$ from the unsigned
weight $A_a$. The convention is
\begin{equation}
 \mathsf A(q,p)=\frac13\sum_{q',p'\in\mathbb Z_3}
 \omega^{pq'-qp'}T(q',p'),\qquad
 \mathsf A(\bm q,\bm p)=\bigotimes_{j=1}^{\ell}\mathsf A(q_j,p_j).
 \label{eq:phasepoint}
\end{equation}
For $D_\ell=3^\ell$, these operators obey
\begin{equation}
 \mathsf A(u)^\dagger=\mathsf A(u),\quad \mathsf A(u)^2=I,\quad
 \Tr\mathsf A(u)=1,\quad
 \Tr[\mathsf A(u)\mathsf A(v)]=D_\ell\delta_{uv},\quad
 \sum_u\mathsf A(u)=D_\ell I.
 \label{eq:phasepointproperties}
\end{equation}
Consequently,
\begin{align}
 W_\rho(u)&=D_\ell^{-1}\Tr[\rho\mathsf A(u)],\qquad
 \sum_u W_\rho(u)=1,\label{eq:wignerdef}\\
 \Tr(\rho\sigma)&=D_\ell\sum_u W_\rho(u)W_\sigma(u).
 \label{eq:wigneroverlap}
\end{align}
Canonical odd-prime phase space supplies stabilizer measurement marginals,
Clifford covariance, and the equivalence between pure state Wigner positivity
and stabilizer states~\cite{supp:Gross2006}. For mixed states, positivity does not
imply a mixture of stabilizer states. A Clifford transformation relabels the
state's cells and its reference support together, preserving the signed and
unsigned support sums below. This operational restriction matters: the split
is not asserted invariant under arbitrary quasiprobability frames, particularly
in even dimension.

The rank one ordered reference and its affine Lagrangian support are
\begin{align}
 |\phi_a^{(\ell)}\rangle&=|a\rangle^{\otimes\ell},\qquad
 W_{\phi_a^{(\ell)}}(\bm q,\bm p)=D_\ell^{-1}\delta_{\bm q,\bm a},
 \label{eq:orderedwigner}\\
 S_a^{(\ell)}&=\{(\bm q,\bm p):\bm q=\bm a\},\qquad
 |S_a^{(\ell)}|=D_\ell,\quad \bm a=(a,\ldots,a).
 \label{eq:returnsupport}
\end{align}
The overlap identity gives the block return directly as a signed support sum:
\begin{equation}
 P_a^{(\ell)}=\langle\phi_a^{(\ell)}|\rho_\ell|\phi_a^{(\ell)}\rangle
 =\sum_{u\in S_a^{(\ell)}}W_{\rho_\ell}(u).
 \label{eq:supp-blockreturn}
\end{equation}
For a pure global state and $\ell=N$, this becomes the squared ordered-state
overlap. A fixed block and the global support probe different information
scales; a singular large-block limit requires its own boundary analysis.

\subsection{Unsigned weight, negative mass, and rate identity}

Separate positive and negative contributions on the same support:
\begin{align}
 P_{a,+}^{(\ell)}&=\sum_{u\in S_a^{(\ell)},\,W(u)>0}W(u),&
 \nu_a^{(\ell)}&=\sum_{u\in S_a^{(\ell)},\,W(u)<0}|W(u)|,\nonumber\\
 P_a^{(\ell)}&=P_{a,+}^{(\ell)}-\nu_a^{(\ell)},&
 A_a^{(\ell)}&=\sum_{u\in S_a^{(\ell)}}|W(u)|=P_{a,+}^{(\ell)}+\nu_a^{(\ell)}.
 \label{eq:unsignedweight}
\end{align}
It follows that
\begin{equation}
 A_a^{(\ell)}-P_a^{(\ell)}=2\nu_a^{(\ell)},\qquad
 0\leq P_a^{(\ell)}\leq A_a^{(\ell)}\leq1.
 \label{eq:canonical-support-bound}
\end{equation}
The upper bound uses both rank one and canonical normalization:
$|\Tr(\rho\mathsf A(u))|\leq\|\mathsf A(u)\|_{\rm op}=1$, and the
support has exactly $D_\ell$ cells, each bounded by $D_\ell^{-1}$.
A disjoint union of $k$ supports has the cell-count bound $A\leq k$;
the single-support bound cannot be transferred to it without further proof.
Negative mass outside $S_a$ does not enter this branch's suppression.
Despite its bound, $A_a$ is a nonlinear auxiliary weight, not a measurement
probability.

On the positive domain $P_a>0$ (which ensures $A_a>0$), define
\begin{align}
 r_a^{(\ell)}&=-\ell^{-1}\ln P_a^{(\ell)},&
 s_a^{(\ell)}&=-\ell^{-1}\ln A_a^{(\ell)},&
 q_a^{(\ell)}&=\ell^{-1}\ln\frac{A_a^{(\ell)}}{P_a^{(\ell)}},
 \label{eq:rates-supp}\\
 r_a^{(\ell)}&=s_a^{(\ell)}+q_a^{(\ell)},&
 s_a^{(\ell)}&\geq0,&0\leq q_a^{(\ell)}&\leq r_a^{(\ell)}.
 \label{eq:rateidentity-supp}
\end{align}
The normalized absolute weights on $S_a$ define a conditional average sign,
\begin{equation}
 \langle\mathrm{sign}\rangle_{a,\ell}
 =\frac{P_a^{(\ell)}}{A_a^{(\ell)}}=e^{-\ell q_a^{(\ell)}}.
 \label{eq:conditionalsign}
\end{equation}
A positive limiting $q_a$ means exponentially strong suppression of the
physical return relative to its unsigned weight. Finite negative mass alone
does not imply an extensive cost. If $P_a=0<A_a$, then $r_a=q_a=+\infty$;
if $A_a=0$, then $P_a=0$, $r_a=s_a=+\infty$, and $q_a$ is undefined.
Finite rate identities and differences are used only where defined.

\subsection{Individual branches, grouping, and physical returns}

For global length $N$, the incoherent ordered-manifold rate is
\begin{equation}
 R_N=-\frac1N\ln\sum_{a=0}^2P_{a,N}
 =-\frac1N\ln\sum_a e^{-Nr_{a,N}},\qquad
 0\leq\min_a r_{a,N}-R_N\leq\frac{\ln3}{N}.
 \label{eq:finite-logsum}
\end{equation}
At finite $N$ it is smooth on a neighborhood with nonzero analytic returns.
If the three limiting rates exist,
\begin{equation}
 R=\min_a r_a.
 \label{eq:laplace-supp}
\end{equation}
A transverse exchange of the minimizing branches produces a cusp; a
crossing preempted by a third lower branch does not.

The grouped comparator is $P_{\rm grp}=P_1+P_2$ and
$A_{\rm grp}=A_1+A_2$. If $P_1=P_2$, then
\begin{equation}
 r_{\rm grp}^{(\ell)}=r_1^{(\ell)}-\frac{\ln2}{\ell}.
 \label{eq:degeneracyshift}
\end{equation}
Only if unsigned symmetry $A_1=A_2$ also holds do we obtain
$s_{\rm grp}^{(\ell)}=s_1^{(\ell)}-\ln2/\ell$ and
$q_{\rm grp}^{(\ell)}=q_1^{(\ell)}$. Both symmetries are independently
checked numerically. The degeneracy shift vanishes in the global limit;
convergence to the same isolated crossing additionally requires uniform
branch control nearby.

Three returns must be distinguished: $P_0$ to the initial state,
$\sum_aP_a$ to the incoherent ordered manifold, and
$3^{-1}|\sum_a\langle\phi_a|\psi\rangle|^2$ to the coherent state
$3^{-1/2}\sum_a|\phi_a\rangle$. The last retains relative phases.
Complete computational-basis dephasing preserves every ordered population
and gives $W(\bm q,\bm p)=p_{\bm q}/3^N\geq0$. Such snapshots does not
follow the original unitary evolution. Ordered returns are global joint
populations, not universal tests of global coherence.

\subsection{Branch ordering, matched crossings, and cusp derivatives}

For individual sectors, set
\begin{equation}
 \Delta r=r_0-r_1,\quad\Delta s=s_0-s_1,\quad\Delta q=q_0-q_1,
 \qquad\Delta r=\Delta s+\Delta q.
 \label{eq:differenceidentity}
\end{equation}
Away from ties, the two comparisons select opposite branches precisely when
\begin{equation}
 \Delta r\,\Delta s<0.
 \label{eq:opposite-ordering-criterion}
\end{equation}
The physical and auxiliary roots are, respectively,
\begin{equation}
 \Delta r(\tau_c)=0,\qquad\Delta s(\tau_s)=0.
 \label{eq:physicalcrossing}
\end{equation}
Sign stripping uses the same state and supports; $W\mapsto|W|$ is generally
neither a physical channel nor a normalized Wigner function. If both supports
are nonnegative throughout the crossing neighborhood, $A_a=P_a$, $q_a=0$,
and the comparisons coincide. Thus distinct isolated roots, matched by
direction of branch exchange in the same time window, suffice to show that signs
alter the branch exchange:
\begin{equation}
 \tau_c\ne\tau_s\quad\Longrightarrow\quad
 \text{return-relevant signs change the matched exchange}.
 \label{eq:sufficientcriterion}
\end{equation}
The converse fails when cancellation affects both branches equally. Nearby
smooth roots admit the first-order interpretation
\begin{equation}
 \tau_c-\tau_s\simeq-
 \frac{\Delta q(\tau_s)}{\partial_\tau\Delta s(\tau_s)+\partial_\tau\Delta q(\tau_s)},
 \label{eq:shiftlinearized}
\end{equation}
provided the denominator is nonzero and higher orders are small. We locate
the roots independently rather than using this approximation. A
thermodynamic displacement requires both limiting roots and evidence of
their separation; $q_1>0$ alone does not establish $\Delta q\ne0$.

\section{Numerical methods}
\label{sec:potts-numerics}

\subsection{Quench and finite open chain evolution}

The three state Potts quench starts from $|0\rangle^{\otimes N}$ at $h_i=0$
and evolves with $h_f=1.5J$ under
\begin{equation}
 H=-J\sum_{j=1}^{N-1}(Z_j^\dagger Z_{j+1}+Z_{j+1}^\dagger Z_j)
   -h\sum_{j=1}^N(X_j+X_j^\dagger).
 \label{eq:potts-ham-supp}
\end{equation}
The infinite chain calculation uses the same local terms and initial state.
We set $J=1$ in the implementation and report $\tau=Jt$. Finite chains
have open boundaries and $N=20,50,75,100$; central blocks are compared
between $N=50$ and 100.

Finite chain time evolving block decimation (TEBD)~\cite{supp:Vidal2004} separates
the onsite field $H_h$ from even and odd diagonal bonds $H_e,H_o$.
The implemented symmetric step is
\begin{equation}
 U_2(\delta t)=e^{-iH_h\delta t/2}e^{-iH_e\delta t/2}
 e^{-iH_o\delta t}e^{-iH_e\delta t/2}e^{-iH_h\delta t/2}
 +O(\delta t^3).
 \label{eq:tebdstep}
\end{equation}
The diagonal bond terms commute. Onsite fields are not distributed into
two-site bond Hamiltonians. After each two-site gate, an SVD discards
singular values below an \emph{absolute} threshold and caps the retained
rank. The global production data use $\delta\tau=0.02$, cap 48, and
cutoff $10^{-10}$; central-block data use cap 32 and cutoff $10^{-9}$.
Time-step comparisons include $\delta\tau=0.04$ and separately evolved
block crossings at $\delta\tau=0.01$.

Norm, accumulated local discarded Schmidt weight, realized bond dimension,
ordered probabilities, local magnetization, and the relative sector-1/2
population difference are recorded. All three overlaps are independently
contracted. Norm preservation does not ensure relative accuracy of
exponentially small returns; summed discarded weights are diagnostics,
not rigorous accumulated state error bounds.

\subsection{Infinite MPS evolution and signed transfer contractions}
\label{sec:imps-method}

The direct thermodynamic calculation uses a two-site infinite MPS Vidal
representation~\cite{supp:Vidal2007Infinite},
$\cdots\Gamma_A\Lambda_A\Gamma_B\Lambda_B\cdots$, with conventional
tensors $A=\Gamma_A\Lambda_A$ and $B=\Gamma_B\Lambda_B$.
The same onsite-half/AB-half/BA-full/AB-half/onsite-half split is used.
Each bond update includes both outer Schmidt spectra, applies the gate,
and truncates by absolute cutoff and rank cap; retained spectra are
normalized.

For a cell of $L=2$ sites, the norm transfer is
$\mathcal E(X)=\sum_{s,t}A^sB^t X(B^t)^\dagger(A^s)^\dagger$.
Positive iterations from identity give left and right environments
$L_e,R_e$, normalized by $\Tr(L_eR_e)=1$. Residuals and an independent
Arnoldi peripheral-spectrum calculation check this iteration. Dividing
each conventional tensor by $\eta^{1/4}$, where $\eta$ is the norm
eigenvalue per cell, gives normalized checkpoint tensors.

Fixing the physical indices to sector $a$ gives the amplitude matrix
$B_a=A^aB^a$. For $k$ cells, the periodic amplitude is $\Tr B_a^k$;
an open amplitude is $l^\dagger B_a^k r$. A simple contributing leading
eigenvalue $\lambda_a$, separated in modulus and with a nonzero boundary
coefficient, yields
\begin{equation}
 r_a=-\frac2L\ln|\lambda_a|+\frac1L\ln\eta.
 \label{eq:signed-transfer-rate}
\end{equation}
The coefficient of a simple eigenvalue in the periodic trace is one.
An open boundary can remove that eigenspace, and equal-modulus
eigenvalues can cancel on subsequences. These cases require retaining
the full contributing spectrum, not blindly taking a spectral radius.
Tests include a pair $+\lambda,-\lambda$ with odd-length cancellation.
Rescaled finite amplitude and norm contractions are compared with spectral
predictions including boundary prefactors. The generic virtual boundaries
used in this test define a finite MPS family, not an independently evolved
open Potts chain.

Signed bulk block probabilities are separately obtained from
\begin{equation}
 P_a^{(2k)}=\frac{\Tr[L_e B_a^kR_e(B_a^\dagger)^k]}{\eta^k},\qquad
 C_a=\frac{(u_a^\dagger L_eu_a)(v_a^\dagger R_ev_a)}{|v_a^\dagger u_a|^2},
 \label{eq:signed-bulk-coefficient}
\end{equation}
where $u_a,v_a$ are right and left eigenvectors. A positive $C_a$ and a
simple dominant modulus yield the same signed exponent as the periodic
global family. This is a conditional statement for the represented MPS,
supported by numerical boundary and finite-contraction checks. It proves
no corresponding equality for unsigned rates.

\subsection{Streamed return supports and distinct limits}

Only $3^\ell$ cells on a specified support are needed, rather than a full
$3^{2\ell}$ phase space array. In the computational basis,
\begin{equation}
 W_a(\bm p)=3^{-\ell}\sum_{\bm x\in\mathbb Z_3^\ell}
 \omega^{-\bm p\cdot\bm x}\rho_\ell(\bm a+2\bm x,\bm a-2\bm x).
 \label{eq:wignerlinefft}
\end{equation}
The finite chain implementation contracts the MPS environments with the
specified ket/bra insertions in batches and absorbs the Fourier factors
into the transfer network. Its streamed cost is approximately
$O(\ell\,3^\ell\chi^3)$, with $O(B_{\rm batch}\chi^2)$ batch memory
in addition to the stored MPS and environments. The infinite state
depth first implementation reuses prefix contractions, storing
$O(\ell\chi^2+3^\ell)$ working/output entries rather than a dense
reduced density matrix.

For the extended block sequence in main-text Fig.~2, we instead split each
support into two halves. Left and right momentum environments
include their local Fourier factors, and their elementwise contraction
gives each complete Wigner cell. Scalar joins are evaluated in bounded
batches; absolute values are taken only after both halves and the physical
boundary environments have been contracted. This changes the contraction
order without dropping cells or modifying the represented state. The
half-support storage scales as $O(3^{\lceil\ell/2\rceil}\chi^2)$ and the
scalar joins as $O(3^\ell\chi^2)$. At $\ell=16$, each sector therefore
includes all $3^{16}=43\,046\,721$ cells. Pointwise comparison with the
original streamed contraction and an independent dense state calculation
checks the new contraction, including odd and even blocks and all sectors.

The projector return, signed Wigner sum, and negative-cell sum are computed
independently. Their comparisons test $\sum W_a=P_a$ and $A_a-P_a=2\nu_a$;
the subsequent identity $r_a=s_a+q_a$ is algebraic postprocessing, not
another independent validation. Taking absolute values of MPS tensors
entry by entry is gauge dependent and destroys interference \emph{inside}
each Wigner cell. It does not compute $\sum|W_a|$.

Global $N\to\infty$ returns and reduced blocks with the order of limits
$\lim_{\ell\to\infty}\lim_{N\to\infty}$ are distinct sequences.
Environmental tracing occurs before the absolute value and cannot
generally be interchanged with it. Agreement of fixed blocks between two
finite environments, or short correlation length, does not establish
equality of these unsigned limits.

\subsection{Collective Ising evolution and probability resolution}
\label{sec:ising-numerics}

For the collective Hamiltonian in Eq.~\eqref{eq:collective-ising-supp},
permutation symmetry confines evolution from $|S,S\rangle$ to the
$S=N/2$ subspace of dimension $N+1$. We diagonalize the Hamiltonian in
the $S_z$ basis and apply its spectral phases to obtain the ordered
return probabilities and $m_z=2\langle S_z\rangle/N$. The selected
crossings are checked independently by tridiagonal spectral evolution
and a sparse matrix exponential acting on the initial state. This
calculation introduces no spatial-boundary or reduced-block approximation.

The sampled crossing selection requires
\begin{equation}
 \min(P_\uparrow,P_\downarrow)\geq\IsingProbabilityFloor
 \label{eq:probabilityfloor}
\end{equation}
at both samples bracketing a zero of $r_\uparrow-r_\downarrow$.
The threshold controls which sampled crossings are selected; it is not
a bound on probability error. For the plotted $N=\IsingNmax$, $h/J=0.7$
comparison, the independently evaluated crossing is
$\tau_c\simeq6.777245970$, with both probabilities approximately
$5.26\times10^{-7}$. This is a resolved late crossing; the selection
criterion does not identify the first physical crossing.

The classical equations~\eqref{eq:classical-eom} are integrated with
DOP853 using relative tolerance $10^{-10}$ and absolute tolerance
$10^{-12}$. The signed magnetization is averaged over
$60\leq\tau\leq100$; at $h/J=1/2$, the exact separatrix
trajectory~\eqref{eq:ising-separatrix-trajectory} supplies this average.

\section{Potts results and numerical validation}
\label{sec:potts-results}

\subsection{Direct infinite MPS calculations}
\label{sec:potts-signed-results}

\begin{figure}[htbp]
 \centering\includegraphics[width=\textwidth]{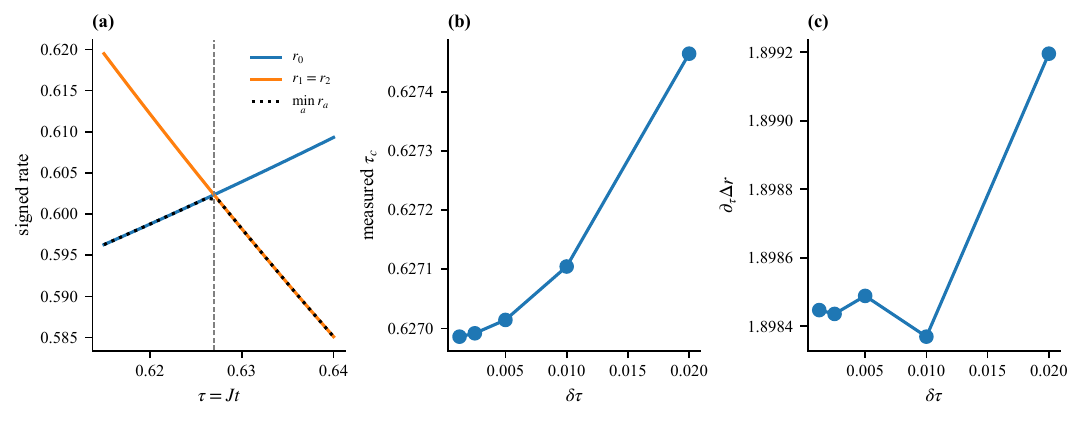}
 \caption{Crossing and cusp stability in the infinite-MPS Potts quench
 from $|0\rangle^{\otimes\infty}$ to $h/J=1.5$.
 (a) Independently contracted return rates and their lower envelope.
 The minimum switches from $r_0$ to $r_1=r_2$, creating a downward cusp.
 (b) Directly recomputed crossing times versus numerical evolution step
 $\delta\tau$. (c) Slope difference
 $\partial_\tau(r_0-r_1)|_{\tau_c}$ at each run's crossing versus the
 same numerical step. It stays close to $1.89845$ as the step is reduced;
 the full variation on the magnified vertical axis is below $0.05\%$.
 The derivative jump of the lower envelope is the negative of this
 quantity. Thus (b) tests the crossing location and (c) tests the
 stability of the cusp's sharpness. These checks concern the signed
 return rates of the computed infinite-chain state.}
 \label{fig:direct-thermo}
\end{figure}

The direct infinite-chain calculation resolves an exchange of the individual
ordered return branches for the $h_i=0\to h_f=1.5J$ quench. A signed
crossing occurs when $\Delta r=r_0-r_1$ changes sign: the initial branch
has the larger return before the crossing, and the competitor afterwards.
We locate the crossing by evolving the state to each trial time used by
the root finder. At the finest evolution step $\delta\tau=\SignedPottsDt$,
\begin{equation}
 \tau_c=\SignedPottsTau,\qquad
 \partial_\tau\Delta r(\tau_c)=\SignedPottsSlope.
 \label{eq:potts-direct-event}
\end{equation}
Sector 2 is contracted independently and agrees with sector 1. The three
rates are approximately $0.6023423275$ at the crossing, with slopes
$r_0'\simeq0.51945036$ and $r_1'=r_2'\simeq-1.37899686$.
The lower envelope $r_{\min}(\tau)=\min_a r_a(\tau)$ therefore switches
from an increasing branch to a decreasing branch, producing the cusp in
Fig.~\ref{fig:direct-thermo}(a). Its derivative jump is
\begin{equation}
 r_{\min}'(\tau_c^+)-r_{\min}'(\tau_c^-)
 =-\partial_\tau\Delta r(\tau_c)\simeq-\SignedPottsSlope.
 \label{eq:potts-cusp-jump}
\end{equation}
Panels (b,c) vary the numerical evolution step $\delta\tau$.
Panel (b) tests the crossing time, while (c) tests the magnitude of the
slope jump in Eq.~\eqref{eq:potts-cusp-jump}, evaluated at each run's
own crossing. The slope difference stays finite as the step decreases;
its variation across the magnified vertical scale is below $0.05\%$.
This supports a cusp with numerically stable sharpness in the computed
infinite MPS. Evolution and contraction checks are given below.

\begin{table}[htbp]
 \centering
 \caption{Direct infinite-MPS signed Potts crossings after the quench from
 $|0\rangle^{\otimes\infty}$ to $h/J=1.5$. The crossing is $r_0=r_1$ with
 an individual competitor; rates refer to the periodic global family of
 the represented two-site MPS under the contributing-spectrum conditions
 of Sec.~\ref{sec:imps-method}. Bond entries are the largest realized
 dimensions on each alternating bond over the run. The absolute Schmidt
 cutoff is $\epsilon$. Displayed digits identify the directly recomputed
 crossings; their accuracy is assessed by varying the step, bond cap, and cutoff.}
 \label{tab:itebd-runs}
\begin{tabular}{cccccc}
\toprule
$\delta\tau$ & cap & realized bonds & $\epsilon$ & $\tau_c$ & $\partial_\tau\Delta r$\\
\midrule
0.02 & 48 & 48/48 & $1.000\times10^{-12}$ & 0.627464382 & 1.8991958\\
0.02 & 128 & 75/79 & $1.000\times10^{-14}$ & 0.627464382 & 1.8991958\\
0.01 & 48 & 48/48 & $1.000\times10^{-12}$ & 0.627104556 & 1.8983697\\
0.01 & 128 & 72/76 & $1.000\times10^{-14}$ & 0.627104556 & 1.8983697\\
0.005 & 48 & 47/48 & $1.000\times10^{-12}$ & 0.627014268 & 1.8984886\\
0.005 & 48 & 48/48 & $1.000\times10^{-14}$ & 0.627014268 & 1.8984886\\
0.005 & 64 & 64/64 & $1.000\times10^{-14}$ & 0.627014268 & 1.8984886\\
0.005 & 96 & 70/72 & $1.000\times10^{-14}$ & 0.627014268 & 1.8984886\\
0.005 & 128 & 58/61 & $1.000\times10^{-13}$ & 0.627014268 & 1.8984886\\
0.005 & 128 & 70/72 & $1.000\times10^{-14}$ & 0.627014268 & 1.8984886\\
0.005 & 128 & 85/88 & $1.000\times10^{-15}$ & 0.627014268 & 1.8984886\\
0.0025 & 48 & 44/48 & $1.000\times10^{-12}$ & 0.626991644 & 1.8984361\\
0.0025 & 128 & 65/70 & $1.000\times10^{-14}$ & 0.626991644 & 1.8984361\\
0.0025 & 128 & 74/79 & $1.000\times10^{-15}$ & 0.626991644 & 1.8984361\\
0.00125 & 128 & 72/74 & $1.000\times10^{-15}$ & 0.626985981 & 1.8984472\\
0.00125 & 256 & 72/74 & $1.000\times10^{-15}$ & 0.626985981 & 1.8984472\\
\bottomrule
\end{tabular}

\end{table}

At $\delta\tau=\SignedPottsPreviousDt$, the crossing is
$\tau_c=\SignedPottsPreviousTau$, a change of $5.66\times10^{-6}$.
The roughly fourfold reduction of successive root shifts is consistent
with second-order evolution, without supplying a rigorous remainder.
At $\delta\tau=0.005$, cutoff $10^{-14}$, caps 48, 64, 96, and 128
realize maximum bonds $48/48$, $64/64$, $70/72$, and $70/72$;
the roots differ by at most $1.22\times10^{-12}$. Cutoff changes from
$10^{-13}$ through $10^{-15}$ at cap 128 give realized bonds
$58/61$, $70/72$, and $85/88$, with root shifts about $2\times10^{-12}$.
At the finest step, caps 128 and 256 both realize $72/74$ over the run
and give identical results; that cap increase does not refine the state.
The state at the crossing has bond dimensions $69/74$.

The last step refinement changes the slope difference by $1.11\times10^{-5}$;
halving its finite-difference increment changes it by $3.01\times10^{-8}$.
At the finest step, the crossing residual $\Delta r(\tau_c)$ is
approximately $-1.32\times10^{-12}$.
The minimum amplitude-transfer modulus gap is $0.743$, condition numbers
are about 3.3, and scaled residuals are at most $1.13\times10^{-14}$.
All tested leading open-boundary coefficients are nonzero. Signed bulk
coefficients are about $0.53365,0.70310,0.70310$; direct contractions
through 128 sites agree with the spectral prediction including prefactors.
Together these checks support an exchange with unequal branch slopes in
the computed infinite-chain state. They provide numerical convergence
evidence, without a rigorous error bound for the exact Hamiltonian.

\FloatBarrier

\subsection{Finite blocks at their own signed and unsigned crossings}
\label{sec:block-own-events}

\begin{figure}[htbp]
 \centering\includegraphics[width=\textwidth]{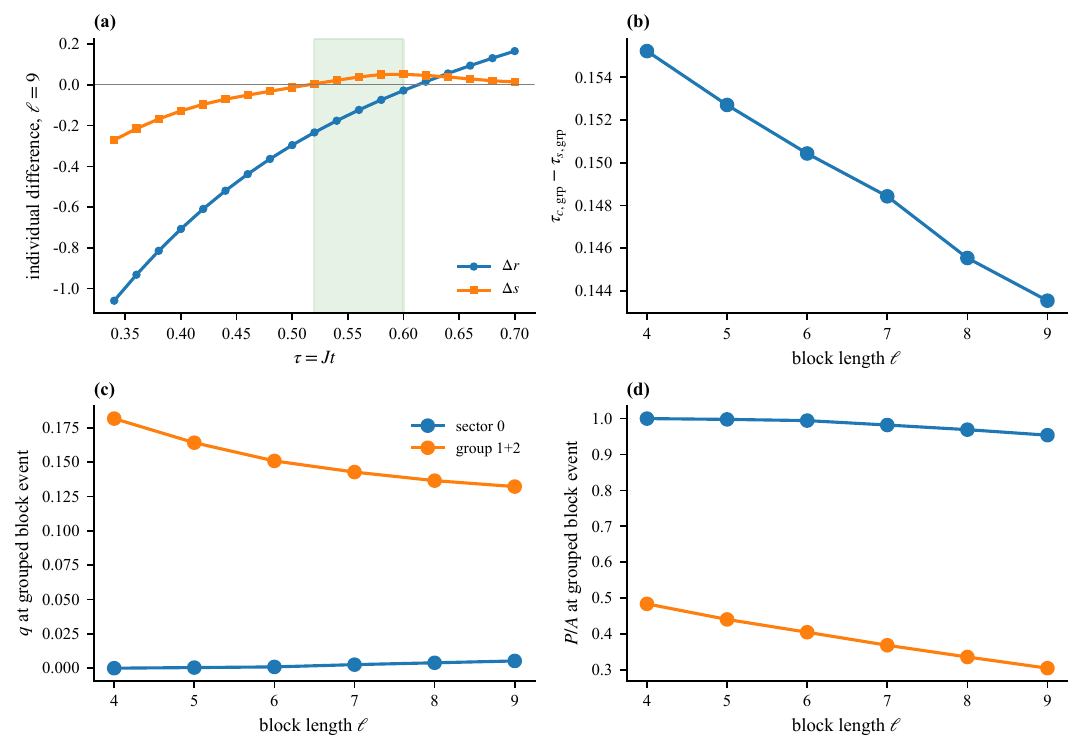}
 \caption{Finite-block Potts interference after the same $h/J=1.5$
 quench, with open-chain production evolution. (a) Time samples
 of individual-sector differences for the central $\ell=9$ block of
 $N=100$; shading marks sampled opposite ordering. Panels (b)--(d)
 instead use the explicitly \emph{grouped} competitor for central blocks
 in $N=100$: signed-minus-unsigned crossing displacement, branch sign costs at the
 physical block crossing, and surviving signed fractions. These panels use
 common interpolation of the sampled probabilities and unsigned weights; Table~\ref{tab:block-nine}
 reports directly recomputed crossings. At each interpolated crossing, the displayed costs and surviving
 fractions are computed from the same interpolated weights. All panels concern finite supports,
 not a thermodynamic unsigned limit.}
 \label{fig:supp-block-sign}
\end{figure}

For central blocks in the finite open chain, direct support sums resolve
different signed and unsigned ordering. For an individual competitor,
the $\ell=9$ block crossings at production step 0.02 are
\begin{equation}
 \tau_c^{(9)}=\BlockIndividualTau,\qquad
 \tau_s^{(9)}=\BlockIndividualUnsignedTau,\qquad
 \Delta s^{(9)}(\tau_c^{(9)})=\BlockIndividualDeltaS.
 \label{eq:block-individual-event}
\end{equation}
The physical tie has $\Delta q=-\Delta s<0$, and sampled times between
the matched roots have opposite signed and unsigned ordering
[Fig.~\ref{fig:supp-block-sign}(a)]. This establishes sign-dependent
branch selection on the specified finite block.

\begin{table}[htbp]
 \centering
 \caption{Directly recomputed central-block crossings for the open Potts
 chain, initially $|0\rangle^{\otimes N}$ and quenched to $h/J=1.5$.
 All rows use $\ell=9$, $\delta\tau=0.02$, cap 32, and absolute cutoff
 $10^{-9}$. Individual sector 1 and grouped sectors $1+2$ are distinct
 comparators. $\tau_c$ and $\tau_s$ are separately recomputed signed and
 unsigned roots with the same direction of branch exchange; $q_{\rm comp}$ is
 evaluated at the signed root. Digits identify the computed crossings; evolution sensitivity is
 assessed separately.}
 \label{tab:block-nine}
\begin{tabular}{clccccc}
\toprule
$N$ & comparison & $\tau_c$ & $\tau_s$ & $\Delta s(\tau_c)$ & $q_{\rm comp}$ & $P_{\rm comp}/A_{\rm comp}$\\
\midrule
50 & individual & 0.612792 & 0.514589 & 0.048811 & 0.085670 & 0.462535\\
50 & grouped & 0.578686 & 0.434310 & 0.127295 & 0.132263 & 0.304110\\
100 & individual & 0.612792 & 0.514589 & 0.048811 & 0.085670 & 0.462535\\
100 & grouped & 0.578686 & 0.434310 & 0.127295 & 0.132263 & 0.304110\\
\bottomrule
\end{tabular}

\end{table}

Grouping changes both finite-block crossing times. At $N=100$, the grouped
physical crossing is $\tau_{c,\mathrm{grp}}=\BlockGroupedTau$, while
$\tau_{s,\mathrm{grp}}=\BlockGroupedUnsignedTau$. Evaluating all weights in the same
state at that physical crossing gives
\begin{align}
 \Delta s_{\mathrm{grp}}&=\BlockGroupedDeltaS,&
 \Delta q_{\mathrm{grp}}&=\BlockGroupedDeltaQ,\label{eq:block-grouped-differences}\\
 q_1=q_{\mathrm{grp}}&=\BlockGroupedCompetingQ,&
 P_1/A_1=P_{\mathrm{grp}}/A_{\mathrm{grp}}&=\BlockGroupedSurvival.
 \label{eq:block-grouped-survival}
\end{align}
The independently contracted sector symmetries justify the equalities.
The primitive values are approximately $P_1=1.627\times10^{-3}$,
$A_1=5.350\times10^{-3}$, and $\nu_1=1.861\times10^{-3}$.
The direct projector and signed Wigner sums agree to $8.7\times10^{-19}$;
the unsigned sector-1/2 relative difference is $5.2\times10^{-15}$.
The survival identity $P_1/A_1=e^{-9q_1}$ is derived from this same record,
whereas the projector and negative-cell checks are independent contractions.

Agreement between $N=50$ and 100 at fixed $\ell$ tests these finite
environments only. A direct $N=50$, $\delta\tau=0.01$ reevaluation gives
the individual crossing $\tau_c=\BlockFineTau$ and
$\Delta s=\BlockFineDeltaS$. The approximately $3.50\times10^{-4}$
time-step shift is much larger than root-solving tolerance. Panels
(b)--(d) of Fig.~\ref{fig:supp-block-sign} compare sector 0 with grouped
sectors $1+2$ across $\ell=4,\ldots,9$, using interpolation of the
sampled probabilities and unsigned weights.

Main-text Fig.~2(b)--(d) extends the grouped comparison to
$\ell=4,\ldots,\FigTwoEllMax$ in independently evolved chains of
$N=50$ and 100. All these block crossings are directly recomputed with
$\delta\tau=0.02$, cap 32, and cutoff $10^{-9}$. Fractional final
evolution steps locate the signed and unsigned roots with time tolerance
$2\times10^{-9}$; the costs and surviving fractions use the same
weights at the signed crossing. At $\ell=16$, the two chains give
$\tau_c\simeq0.598278223$, $\tau_s\simeq0.465619053$,
$\tau_c-\tau_s=\FigTwoShiftHighN$, and
$q_\perp(\tau_c)=\FigTwoQComp$. The surviving fraction is
$P_\perp/A_\perp=\FigTwoSurvivalComp$. Across the full sequence, the
largest chain-size differences in the displacement and cost are below
$1.1\times10^{-10}$ and $2.0\times10^{-10}$, respectively.
The independent projector/Wigner reconstruction error is below
$1.7\times10^{-16}$, and the largest crossing residual is below
$6.6\times10^{-10}$. These contraction and root checks are distinct from
the evolution sensitivity in Table~\ref{tab:fig2-extension-refinement}.
Halving the step changes the sixteen-site displacement by
$3.20\times10^{-4}$ and the competing cost by $1.29\times10^{-4}$;
both remain positive. The simultaneous cap increase and tighter cutoff
change these quantities by $2.15\times10^{-9}$ and
$4.02\times10^{-9}$. The main-text fit ranges use this directly
recomputed sequence for
$\ell=4,\ldots,16$, with linear and quadratic fits, lower cutoffs
$\ell=4,5,6$, and both chain sizes. The resulting spread measures
sensitivity to the fit choice. Positive extrapolated intercepts
do not establish the unsigned thermodynamic limit.

\begin{table}[htbp]
 \centering
 \caption{Evolution refinement for the sixteen-site central block
 in the $N=50$ open chain. All rows compare sector 0 with the independently
 contracted grouped competitor $1+2$, using direct evolved-state signed
 and unsigned crossings. The second row halves the evolution step; the third
 jointly raises the cap and tightens the absolute Schmidt cutoff
 $\epsilon$. Realized maximum bonds at the signed crossing are 29, 28, and
 36, respectively. Digits identify the directly recomputed crossings; differences between
 rows quantify the tested evolution sensitivity.}
 \label{tab:fig2-extension-refinement}
\begin{tabular}{ccccccc}
\toprule
$\delta\tau$ & $\chi_{\max}$ & $\epsilon$ & $\tau_c$ & $\tau_s$ & $\tau_c-\tau_s$ & $q_\perp(\tau_c)$ \\
\midrule
0.02 & 32 & $10^{-9}$ & 0.598278223 & 0.465619053 & 0.132659170 & 0.117158519 \\
0.01 & 32 & $10^{-9}$ & 0.597922489 & 0.464943199 & 0.132979290 & 0.117287613 \\
0.02 & 48 & $10^{-10}$ & 0.598278224 & 0.465619051 & 0.132659173 & 0.117158523 \\
\bottomrule
\end{tabular}

\end{table}

\FloatBarrier

\section{Collective Ising comparison}
\label{sec:information}\label{sec:ising-benchmark}

\subsection{Collective dynamics and the classical separatrix}

The collective Ising Hamiltonian is
\begin{equation}
 H=-\frac{2J}{N}S_z^2-2hS_x,\qquad S=N/2.
 \label{eq:collective-ising-supp}
\end{equation}
The north-pole state $|S,S\rangle$ remains in the permutation-symmetric
subspace of dimension $N+1$; numerical methods and probability resolution
are specified in Sec.~\ref{sec:ising-numerics}. The ordered returns are
\begin{equation}
 P_\uparrow=|\langle S,S|\psi(t)\rangle|^2,\quad
 P_\downarrow=|\langle S,-S|\psi(t)\rangle|^2,\qquad
 r_{\uparrow,\downarrow}=-N^{-1}\ln P_{\uparrow,\downarrow}.
 \label{eq:ising-returns}
\end{equation}
Their incoherent sum gives a smooth finite-size rate; its limiting lower
envelope can develop a cusp.

For the classical unit spin $\bm m=2\langle\bm S\rangle/N$,
\begin{equation}
 e(\bm m)=-\frac J2m_z^2-hm_x,
 \label{eq:ising-classical-energy}
\end{equation}
and precession gives
\begin{equation}
 \dot m_x=2Jm_ym_z,\qquad
 \dot m_y=2hm_z-2Jm_xm_z,\qquad
 \dot m_z=-2hm_y.
 \label{eq:classical-eom}
\end{equation}
The initial energy is $-J/2$ and the saddle at $(m_x,m_z)=(1,0)$ has
energy $-h$. Their equality fixes the separatrix
\begin{equation}
 h_c/J=1/2.
 \label{eq:ising-hc}
\end{equation}
The signed late-time average
\begin{equation}
 \overline m_z=(T_2-T_1)^{-1}\int_{T_1}^{T_2}m_z(t)\,dt
 \label{eq:signedmag}
\end{equation}
tests dynamical order; replacing it by $\overline{|m_z|}$ can mistake
oscillations between symmetry-related sectors for persistent order.
At the critical field the exact trajectory is
\begin{equation}
 m_x(\tau)=\tanh^2\tau,\quad
 m_y(\tau)=\operatorname{sech}\tau\tanh\tau,\quad
 m_z(\tau)=\operatorname{sech}\tau.
 \label{eq:ising-separatrix-trajectory}
\end{equation}
Its signed mean over $\tau\in[60,100]$ is approximately
$4.38\times10^{-28}$, which supplies the critical-field value in the scan.

\subsection{Local temporal zeros and global return competition}

\begin{figure}[htbp]
 \centering\includegraphics[width=\textwidth]{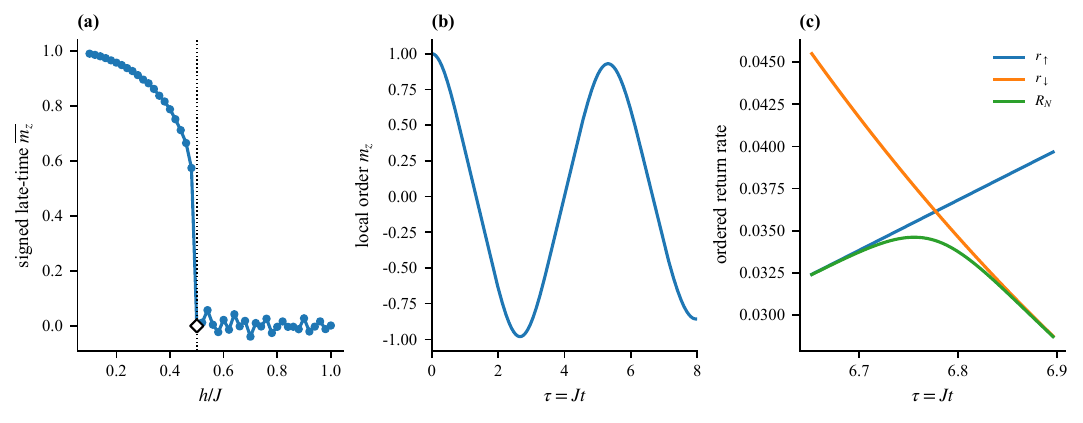}
 \caption{Information scales in the collective Ising model, initially
 at the north pole. (a) Signed late-time magnetization versus field;
 the critical mean-field sample uses the analytic separatrix
 Eq.~\eqref{eq:ising-separatrix-trajectory}. (b) Along $N=400$, $h/J=0.7$,
 local magnetization has a smooth temporal zero distinct from the
 selected return exchange. (c) Global ordered rates and their smooth
 finite-size combination at that exchange. The field-driven local
 diagnostic and the temporal branch crossing are different questions;
 no universal identification of DQPT-I and DQPT-II is inferred.}
 \label{fig:supp-ising-benchmark}
\end{figure}

Figure~\ref{fig:supp-ising-benchmark} compares field-dependent local
dynamical order with global branch competition. Along the $N=400$,
$h/J=0.7$ trajectory, the local magnetization crosses zero near
$\tau=6.650$ and stays smooth at the selected return exchange near
$6.777$, where $m_z\simeq-0.122$. Equal probabilities for the two
extreme spin configurations need not balance all positive and negative
spin projections contributing to $m_z$. A local temporal zero therefore
cannot replace the return-branch condition. Its finite-system time offset
does not establish a separation of thermodynamic critical points, and a
temporal zero is not the field-driven DQPT-I at $h/J=1/2$.

Qubits lack the same combination of odd-prime canonical positivity and
covariance used in Sec.~\ref{sec:qutritphase}; different valid frames can
assign different negativity. The Ising comparison therefore uses
representation-independent probabilities and magnetization.

\FloatBarrier


\begin{thebibliography}{27}%
\makeatletter
\providecommand \@ifxundefined [1]{%
 \@ifx{#1\undefined}
}%
\providecommand \@ifnum [1]{%
 \ifnum #1\expandafter \@firstoftwo
 \else \expandafter \@secondoftwo
 \fi
}%
\providecommand \@ifx [1]{%
 \ifx #1\expandafter \@firstoftwo
 \else \expandafter \@secondoftwo
 \fi
}%
\providecommand \natexlab [1]{#1}%
\providecommand \enquote  [1]{``#1''}%
\providecommand \bibnamefont  [1]{#1}%
\providecommand \bibfnamefont [1]{#1}%
\providecommand \citenamefont [1]{#1}%
\providecommand \href@noop [0]{\@secondoftwo}%
\providecommand \href [0]{\begingroup \@sanitize@url \@href}%
\providecommand \@href[1]{\@@startlink{#1}\@@href}%
\providecommand \@@href[1]{\endgroup#1\@@endlink}%
\providecommand \@sanitize@url [0]{\catcode `\\12\catcode `\$12\catcode `\&12\catcode `\#12\catcode `\^12\catcode `\_12\catcode `\%12\relax}%
\providecommand \@@startlink[1]{}%
\providecommand \@@endlink[0]{}%
\providecommand \url  [0]{\begingroup\@sanitize@url \@url }%
\providecommand \@url [1]{\endgroup\@href {#1}{\urlprefix }}%
\providecommand \urlprefix  [0]{URL }%
\providecommand \Eprint [0]{\href }%
\providecommand \doibase [0]{https://doi.org/}%
\providecommand \selectlanguage [0]{\@gobble}%
\providecommand \bibinfo  [0]{\@secondoftwo}%
\providecommand \bibfield  [0]{\@secondoftwo}%
\providecommand \translation [1]{[#1]}%
\providecommand \BibitemOpen [0]{}%
\providecommand \bibitemStop [0]{}%
\providecommand \bibitemNoStop [0]{.\EOS\space}%
\providecommand \EOS [0]{\spacefactor3000\relax}%
\providecommand \BibitemShut  [1]{\csname bibitem#1\endcsname}%
\let\auto@bib@innerbib\@empty
\bibitem [{\citenamefont {Sciolla}\ and\ \citenamefont {Biroli}(2010)}]{Sciolla2010}%
  \BibitemOpen
  \bibfield  {author} {\bibinfo {author} {\bibfnamefont {B.}~\bibnamefont {Sciolla}}\ and\ \bibinfo {author} {\bibfnamefont {G.}~\bibnamefont {Biroli}},\ }\bibfield  {title} {\bibinfo {title} {Quantum quenches and off-equilibrium dynamical transition in the infinite-dimensional {Bose-Hubbard} model},\ }\href {https://doi.org/10.1103/PhysRevLett.105.220401} {\bibfield  {journal} {\bibinfo  {journal} {Physical Review Letters}\ }\textbf {\bibinfo {volume} {105}},\ \bibinfo {pages} {220401} (\bibinfo {year} {2010})}\BibitemShut {NoStop}%
\bibitem [{\citenamefont {Heyl}\ \emph {et~al.}(2013)\citenamefont {Heyl}, \citenamefont {Polkovnikov},\ and\ \citenamefont {Kehrein}}]{Heyl2013}%
  \BibitemOpen
  \bibfield  {author} {\bibinfo {author} {\bibfnamefont {M.}~\bibnamefont {Heyl}}, \bibinfo {author} {\bibfnamefont {A.}~\bibnamefont {Polkovnikov}},\ and\ \bibinfo {author} {\bibfnamefont {S.}~\bibnamefont {Kehrein}},\ }\bibfield  {title} {\bibinfo {title} {Dynamical quantum phase transitions in the transverse-field {Ising} model},\ }\href {https://doi.org/10.1103/PhysRevLett.110.135704} {\bibfield  {journal} {\bibinfo  {journal} {Physical Review Letters}\ }\textbf {\bibinfo {volume} {110}},\ \bibinfo {pages} {135704} (\bibinfo {year} {2013})}\BibitemShut {NoStop}%
\bibitem [{\citenamefont {Heyl}(2018)}]{Heyl2018}%
  \BibitemOpen
  \bibfield  {author} {\bibinfo {author} {\bibfnamefont {M.}~\bibnamefont {Heyl}},\ }\bibfield  {title} {\bibinfo {title} {Dynamical quantum phase transitions: a review},\ }\href {https://doi.org/10.1088/1361-6633/aaaf9a} {\bibfield  {journal} {\bibinfo  {journal} {Reports on Progress in Physics}\ }\textbf {\bibinfo {volume} {81}},\ \bibinfo {pages} {054001} (\bibinfo {year} {2018})}\BibitemShut {NoStop}%
\bibitem [{\citenamefont {Jurcevic}\ \emph {et~al.}(2017)\citenamefont {Jurcevic}, \citenamefont {Shen}, \citenamefont {Hauke}, \citenamefont {Maier}, \citenamefont {Brydges}, \citenamefont {Hempel}, \citenamefont {Lanyon}, \citenamefont {Heyl}, \citenamefont {Blatt},\ and\ \citenamefont {Roos}}]{Jurcevic2017}%
  \BibitemOpen
  \bibfield  {author} {\bibinfo {author} {\bibfnamefont {P.}~\bibnamefont {Jurcevic}}, \bibinfo {author} {\bibfnamefont {H.}~\bibnamefont {Shen}}, \bibinfo {author} {\bibfnamefont {P.}~\bibnamefont {Hauke}}, \bibinfo {author} {\bibfnamefont {C.}~\bibnamefont {Maier}}, \bibinfo {author} {\bibfnamefont {T.}~\bibnamefont {Brydges}}, \bibinfo {author} {\bibfnamefont {C.}~\bibnamefont {Hempel}}, \bibinfo {author} {\bibfnamefont {B.~P.}\ \bibnamefont {Lanyon}}, \bibinfo {author} {\bibfnamefont {M.}~\bibnamefont {Heyl}}, \bibinfo {author} {\bibfnamefont {R.}~\bibnamefont {Blatt}},\ and\ \bibinfo {author} {\bibfnamefont {C.~F.}\ \bibnamefont {Roos}},\ }\bibfield  {title} {\bibinfo {title} {Direct observation of dynamical quantum phase transitions in an interacting many-body system},\ }\href {https://doi.org/10.1103/PhysRevLett.119.080501} {\bibfield  {journal} {\bibinfo  {journal} {Physical Review Letters}\ }\textbf {\bibinfo {volume} {119}},\ \bibinfo {pages} {080501} (\bibinfo {year} {2017})}\BibitemShut {NoStop}%
\bibitem [{\citenamefont {Zhang}\ \emph {et~al.}(2017)\citenamefont {Zhang}, \citenamefont {Pagano}, \citenamefont {Hess}, \citenamefont {Kyprianidis}, \citenamefont {Becker}, \citenamefont {Kaplan}, \citenamefont {Gorshkov}, \citenamefont {Gong},\ and\ \citenamefont {Monroe}}]{Zhang2017}%
  \BibitemOpen
  \bibfield  {author} {\bibinfo {author} {\bibfnamefont {J.}~\bibnamefont {Zhang}}, \bibinfo {author} {\bibfnamefont {G.}~\bibnamefont {Pagano}}, \bibinfo {author} {\bibfnamefont {P.~W.}\ \bibnamefont {Hess}}, \bibinfo {author} {\bibfnamefont {A.}~\bibnamefont {Kyprianidis}}, \bibinfo {author} {\bibfnamefont {P.}~\bibnamefont {Becker}}, \bibinfo {author} {\bibfnamefont {H.}~\bibnamefont {Kaplan}}, \bibinfo {author} {\bibfnamefont {A.~V.}\ \bibnamefont {Gorshkov}}, \bibinfo {author} {\bibfnamefont {Z.-X.}\ \bibnamefont {Gong}},\ and\ \bibinfo {author} {\bibfnamefont {C.}~\bibnamefont {Monroe}},\ }\bibfield  {title} {\bibinfo {title} {Observation of a many-body dynamical phase transition with a 53-qubit quantum simulator},\ }\href {https://doi.org/10.1038/nature24654} {\bibfield  {journal} {\bibinfo  {journal} {Nature}\ }\textbf {\bibinfo {volume} {551}},\ \bibinfo {pages} {601} (\bibinfo {year} {2017})}\BibitemShut {NoStop}%
\bibitem [{\citenamefont {\v{Z}unkovi\v{c}}\ \emph {et~al.}(2018)\citenamefont {\v{Z}unkovi\v{c}}, \citenamefont {Heyl}, \citenamefont {Knap},\ and\ \citenamefont {Silva}}]{Zunkovic2018}%
  \BibitemOpen
  \bibfield  {author} {\bibinfo {author} {\bibfnamefont {B.}~\bibnamefont {\v{Z}unkovi\v{c}}}, \bibinfo {author} {\bibfnamefont {M.}~\bibnamefont {Heyl}}, \bibinfo {author} {\bibfnamefont {M.}~\bibnamefont {Knap}},\ and\ \bibinfo {author} {\bibfnamefont {A.}~\bibnamefont {Silva}},\ }\bibfield  {title} {\bibinfo {title} {Dynamical quantum phase transitions in spin chains with long-range interactions: Merging different concepts of nonequilibrium criticality},\ }\href {https://doi.org/10.1103/PhysRevLett.120.130601} {\bibfield  {journal} {\bibinfo  {journal} {Physical Review Letters}\ }\textbf {\bibinfo {volume} {120}},\ \bibinfo {pages} {130601} (\bibinfo {year} {2018})}\BibitemShut {NoStop}%
\bibitem [{\citenamefont {Corps}\ and\ \citenamefont {Rela\~no}(2023)}]{Corps2023}%
  \BibitemOpen
  \bibfield  {author} {\bibinfo {author} {\bibfnamefont {{\'A}.~L.}\ \bibnamefont {Corps}}\ and\ \bibinfo {author} {\bibfnamefont {A.}~\bibnamefont {Rela\~no}},\ }\bibfield  {title} {\bibinfo {title} {Theory of dynamical phase transitions in quantum systems with symmetry-breaking eigenstates},\ }\href {https://doi.org/10.1103/PhysRevLett.130.100402} {\bibfield  {journal} {\bibinfo  {journal} {Physical Review Letters}\ }\textbf {\bibinfo {volume} {130}},\ \bibinfo {pages} {100402} (\bibinfo {year} {2023})}\BibitemShut {NoStop}%
\bibitem [{\citenamefont {Van~Damme}\ \emph {et~al.}(2023)\citenamefont {Van~Damme}, \citenamefont {Desaules}, \citenamefont {Papi\ifmmode~\acute{c}\else \'{c}\fi{}},\ and\ \citenamefont {Halimeh}}]{PRR_2023_Halimeh}%
  \BibitemOpen
  \bibfield  {author} {\bibinfo {author} {\bibfnamefont {M.}~\bibnamefont {Van~Damme}}, \bibinfo {author} {\bibfnamefont {J.-Y.}\ \bibnamefont {Desaules}}, \bibinfo {author} {\bibfnamefont {Z.}~\bibnamefont {Papi\ifmmode~\acute{c}\else \'{c}\fi{}}},\ and\ \bibinfo {author} {\bibfnamefont {J.~C.}\ \bibnamefont {Halimeh}},\ }\bibfield  {title} {\bibinfo {title} {Anatomy of dynamical quantum phase transitions},\ }\href {https://doi.org/10.1103/PhysRevResearch.5.033090} {\bibfield  {journal} {\bibinfo  {journal} {Phys. Rev. Res.}\ }\textbf {\bibinfo {volume} {5}},\ \bibinfo {pages} {033090} (\bibinfo {year} {2023})}\BibitemShut {NoStop}%
\bibitem [{\citenamefont {Bandyopadhyay}\ \emph {et~al.}(2021)\citenamefont {Bandyopadhyay}, \citenamefont {Polkovnikov},\ and\ \citenamefont {Dutta}}]{PRL_2021_Dutta}%
  \BibitemOpen
  \bibfield  {author} {\bibinfo {author} {\bibfnamefont {S.}~\bibnamefont {Bandyopadhyay}}, \bibinfo {author} {\bibfnamefont {A.}~\bibnamefont {Polkovnikov}},\ and\ \bibinfo {author} {\bibfnamefont {A.}~\bibnamefont {Dutta}},\ }\bibfield  {title} {\bibinfo {title} {Observing dynamical quantum phase transitions through quasilocal string operators},\ }\href {https://doi.org/10.1103/PhysRevLett.126.200602} {\bibfield  {journal} {\bibinfo  {journal} {Phys. Rev. Lett.}\ }\textbf {\bibinfo {volume} {126}},\ \bibinfo {pages} {200602} (\bibinfo {year} {2021})}\BibitemShut {NoStop}%
\bibitem [{\citenamefont {De~Nicola}\ \emph {et~al.}(2021)\citenamefont {De~Nicola}, \citenamefont {Michailidis},\ and\ \citenamefont {Serbyn}}]{PRL_2021_Maksym}%
  \BibitemOpen
  \bibfield  {author} {\bibinfo {author} {\bibfnamefont {S.}~\bibnamefont {De~Nicola}}, \bibinfo {author} {\bibfnamefont {A.~A.}\ \bibnamefont {Michailidis}},\ and\ \bibinfo {author} {\bibfnamefont {M.}~\bibnamefont {Serbyn}},\ }\bibfield  {title} {\bibinfo {title} {Entanglement view of dynamical quantum phase transitions},\ }\href {https://doi.org/10.1103/PhysRevLett.126.040602} {\bibfield  {journal} {\bibinfo  {journal} {Phys. Rev. Lett.}\ }\textbf {\bibinfo {volume} {126}},\ \bibinfo {pages} {040602} (\bibinfo {year} {2021})}\BibitemShut {NoStop}%
\bibitem [{\citenamefont {Wootters}(1987)}]{Wootters1987}%
  \BibitemOpen
  \bibfield  {author} {\bibinfo {author} {\bibfnamefont {W.~K.}\ \bibnamefont {Wootters}},\ }\bibfield  {title} {\bibinfo {title} {A {Wigner}-function formulation of finite-state quantum mechanics},\ }\href {https://doi.org/10.1016/0003-4916(87)90176-X} {\bibfield  {journal} {\bibinfo  {journal} {Annals of Physics}\ }\textbf {\bibinfo {volume} {176}},\ \bibinfo {pages} {1} (\bibinfo {year} {1987})}\BibitemShut {NoStop}%
\bibitem [{\citenamefont {Mzaouali}\ \emph {et~al.}(2019)\citenamefont {Mzaouali}, \citenamefont {Campbell},\ and\ \citenamefont {El~Baz}}]{Mzaouali2019}%
  \BibitemOpen
  \bibfield  {author} {\bibinfo {author} {\bibfnamefont {Z.}~\bibnamefont {Mzaouali}}, \bibinfo {author} {\bibfnamefont {S.}~\bibnamefont {Campbell}},\ and\ \bibinfo {author} {\bibfnamefont {M.}~\bibnamefont {El~Baz}},\ }\bibfield  {title} {\bibinfo {title} {Discrete and generalized phase space techniques in critical quantum spin chains},\ }\href {https://doi.org/10.1016/j.physleta.2019.125932} {\bibfield  {journal} {\bibinfo  {journal} {Physics Letters A}\ }\textbf {\bibinfo {volume} {383}},\ \bibinfo {pages} {125932} (\bibinfo {year} {2019})}\BibitemShut {NoStop}%
\bibitem [{\citenamefont {Hawary}\ \emph {et~al.}(2024)\citenamefont {Hawary}, \citenamefont {Azzouz}, \citenamefont {Baz}, \citenamefont {Deffner}, \citenamefont {Gardas},\ and\ \citenamefont {Mzaouali}}]{PRE_2024_Hawary}%
  \BibitemOpen
  \bibfield  {author} {\bibinfo {author} {\bibfnamefont {K.~E.}\ \bibnamefont {Hawary}}, \bibinfo {author} {\bibfnamefont {M.}~\bibnamefont {Azzouz}}, \bibinfo {author} {\bibfnamefont {M.~E.}\ \bibnamefont {Baz}}, \bibinfo {author} {\bibfnamefont {S.}~\bibnamefont {Deffner}}, \bibinfo {author} {\bibfnamefont {B.}~\bibnamefont {Gardas}},\ and\ \bibinfo {author} {\bibfnamefont {Z.}~\bibnamefont {Mzaouali}},\ }\bibfield  {title} {\bibinfo {title} {Navigating the phase diagram of quantum many-body systems in phase space},\ }\href {https://doi.org/10.1103/PhysRevE.110.014120} {\bibfield  {journal} {\bibinfo  {journal} {Phys. Rev. E}\ }\textbf {\bibinfo {volume} {110}},\ \bibinfo {pages} {014120} (\bibinfo {year} {2024})}\BibitemShut {NoStop}%
\bibitem [{\citenamefont {White}\ \emph {et~al.}(2021)\citenamefont {White}, \citenamefont {Cao},\ and\ \citenamefont {Swingle}}]{White2021}%
  \BibitemOpen
  \bibfield  {author} {\bibinfo {author} {\bibfnamefont {C.~D.}\ \bibnamefont {White}}, \bibinfo {author} {\bibfnamefont {C.}~\bibnamefont {Cao}},\ and\ \bibinfo {author} {\bibfnamefont {B.}~\bibnamefont {Swingle}},\ }\bibfield  {title} {\bibinfo {title} {Conformal field theories are magical},\ }\href {https://doi.org/10.1103/PhysRevB.103.075145} {\bibfield  {journal} {\bibinfo  {journal} {Physical Review B}\ }\textbf {\bibinfo {volume} {103}},\ \bibinfo {pages} {075145} (\bibinfo {year} {2021})}\BibitemShut {NoStop}%
\bibitem [{\citenamefont {Tarabunga}(2024)}]{Tarabunga2024}%
  \BibitemOpen
  \bibfield  {author} {\bibinfo {author} {\bibfnamefont {P.~S.}\ \bibnamefont {Tarabunga}},\ }\bibfield  {title} {\bibinfo {title} {Critical behaviors of non-stabilizerness in quantum spin chains},\ }\href {https://doi.org/10.22331/q-2024-07-17-1413} {\bibfield  {journal} {\bibinfo  {journal} {Quantum}\ }\textbf {\bibinfo {volume} {8}},\ \bibinfo {pages} {1413} (\bibinfo {year} {2024})}\BibitemShut {NoStop}%
\bibitem [{\citenamefont {Karrasch}\ and\ \citenamefont {Schuricht}(2017)}]{Karrasch2017}%
  \BibitemOpen
  \bibfield  {author} {\bibinfo {author} {\bibfnamefont {C.}~\bibnamefont {Karrasch}}\ and\ \bibinfo {author} {\bibfnamefont {D.}~\bibnamefont {Schuricht}},\ }\bibfield  {title} {\bibinfo {title} {Dynamical quantum phase transitions in the quantum {Potts} chain},\ }\href {https://doi.org/10.1103/PhysRevB.95.075143} {\bibfield  {journal} {\bibinfo  {journal} {Physical Review B}\ }\textbf {\bibinfo {volume} {95}},\ \bibinfo {pages} {075143} (\bibinfo {year} {2017})}\BibitemShut {NoStop}%
\bibitem [{\citenamefont {Khasseh}\ \emph {et~al.}(2020)\citenamefont {Khasseh}, \citenamefont {Russomanno}, \citenamefont {Schmitt}, \citenamefont {Heyl},\ and\ \citenamefont {Fazio}}]{Khasseh2020}%
  \BibitemOpen
  \bibfield  {author} {\bibinfo {author} {\bibfnamefont {R.}~\bibnamefont {Khasseh}}, \bibinfo {author} {\bibfnamefont {A.}~\bibnamefont {Russomanno}}, \bibinfo {author} {\bibfnamefont {M.}~\bibnamefont {Schmitt}}, \bibinfo {author} {\bibfnamefont {M.}~\bibnamefont {Heyl}},\ and\ \bibinfo {author} {\bibfnamefont {R.}~\bibnamefont {Fazio}},\ }\bibfield  {title} {\bibinfo {title} {Discrete truncated {Wigner} approach to dynamical phase transitions in {Ising} models after a quantum quench},\ }\href {https://doi.org/10.1103/PhysRevB.102.014303} {\bibfield  {journal} {\bibinfo  {journal} {Physical Review B}\ }\textbf {\bibinfo {volume} {102}},\ \bibinfo {pages} {014303} (\bibinfo {year} {2020})}\BibitemShut {NoStop}%
\bibitem [{\citenamefont {Veitch}\ \emph {et~al.}(2014)\citenamefont {Veitch}, \citenamefont {Mousavian}, \citenamefont {Gottesman},\ and\ \citenamefont {Emerson}}]{Veitch2014}%
  \BibitemOpen
  \bibfield  {author} {\bibinfo {author} {\bibfnamefont {V.}~\bibnamefont {Veitch}}, \bibinfo {author} {\bibfnamefont {S.~A.~H.}\ \bibnamefont {Mousavian}}, \bibinfo {author} {\bibfnamefont {D.}~\bibnamefont {Gottesman}},\ and\ \bibinfo {author} {\bibfnamefont {J.}~\bibnamefont {Emerson}},\ }\bibfield  {title} {\bibinfo {title} {The resource theory of stabilizer quantum computation},\ }\href {https://doi.org/10.1088/1367-2630/16/1/013009} {\bibfield  {journal} {\bibinfo  {journal} {New Journal of Physics}\ }\textbf {\bibinfo {volume} {16}},\ \bibinfo {pages} {013009} (\bibinfo {year} {2014})}\BibitemShut {NoStop}%
\bibitem [{\citenamefont {Pashayan}\ \emph {et~al.}(2015)\citenamefont {Pashayan}, \citenamefont {Wallman},\ and\ \citenamefont {Bartlett}}]{Pashayan2015}%
  \BibitemOpen
  \bibfield  {author} {\bibinfo {author} {\bibfnamefont {H.}~\bibnamefont {Pashayan}}, \bibinfo {author} {\bibfnamefont {J.~J.}\ \bibnamefont {Wallman}},\ and\ \bibinfo {author} {\bibfnamefont {S.~D.}\ \bibnamefont {Bartlett}},\ }\bibfield  {title} {\bibinfo {title} {Estimating outcome probabilities of quantum circuits using quasiprobabilities},\ }\href {https://doi.org/10.1103/PhysRevLett.115.070501} {\bibfield  {journal} {\bibinfo  {journal} {Physical Review Letters}\ }\textbf {\bibinfo {volume} {115}},\ \bibinfo {pages} {070501} (\bibinfo {year} {2015})}\BibitemShut {NoStop}%
\bibitem [{\citenamefont {Gross}(2006)}]{Gross2006}%
  \BibitemOpen
  \bibfield  {author} {\bibinfo {author} {\bibfnamefont {D.}~\bibnamefont {Gross}},\ }\bibfield  {title} {\bibinfo {title} {Hudson's theorem for finite-dimensional quantum systems},\ }\href {https://doi.org/10.1063/1.2393152} {\bibfield  {journal} {\bibinfo  {journal} {Journal of Mathematical Physics}\ }\textbf {\bibinfo {volume} {47}},\ \bibinfo {pages} {122107} (\bibinfo {year} {2006})}\BibitemShut {NoStop}%
\bibitem [{\citenamefont {Zhu}(2016)}]{Zhu2016}%
  \BibitemOpen
  \bibfield  {author} {\bibinfo {author} {\bibfnamefont {H.}~\bibnamefont {Zhu}},\ }\bibfield  {title} {\bibinfo {title} {Permutation symmetry determines the discrete {Wigner} function},\ }\href {https://doi.org/10.1103/PhysRevLett.116.040501} {\bibfield  {journal} {\bibinfo  {journal} {Physical Review Letters}\ }\textbf {\bibinfo {volume} {116}},\ \bibinfo {pages} {040501} (\bibinfo {year} {2016})}\BibitemShut {NoStop}%
\bibitem [{Sup()}]{Supplement}%
  \BibitemOpen
  \href@noop {} {}\bibinfo {note} {See Supplemental Material for proofs, algorithms, convergence tests, data tables, and additional controls.}\BibitemShut {Stop}%
\bibitem [{\citenamefont {Vidal}(2004)}]{Vidal2004}%
  \BibitemOpen
  \bibfield  {author} {\bibinfo {author} {\bibfnamefont {G.}~\bibnamefont {Vidal}},\ }\bibfield  {title} {\bibinfo {title} {Efficient simulation of one-dimensional quantum many-body systems},\ }\href {https://doi.org/10.1103/PhysRevLett.93.040502} {\bibfield  {journal} {\bibinfo  {journal} {Physical Review Letters}\ }\textbf {\bibinfo {volume} {93}},\ \bibinfo {pages} {040502} (\bibinfo {year} {2004})}\BibitemShut {NoStop}%
\bibitem [{\citenamefont {Suzuki}(1976)}]{Suzuki1976}%
  \BibitemOpen
  \bibfield  {author} {\bibinfo {author} {\bibfnamefont {M.}~\bibnamefont {Suzuki}},\ }\bibfield  {title} {\bibinfo {title} {Generalized trotter's formula and systematic approximants of exponential operators and inner derivations with applications to many-body problems},\ }\href {https://doi.org/10.1007/BF01609348} {\bibfield  {journal} {\bibinfo  {journal} {Communications in Mathematical Physics}\ }\textbf {\bibinfo {volume} {51}},\ \bibinfo {pages} {183} (\bibinfo {year} {1976})}\BibitemShut {NoStop}%
\bibitem [{\citenamefont {Halimeh}\ \emph {et~al.}(2021)\citenamefont {Halimeh}, \citenamefont {Trapin}, \citenamefont {Van~Damme},\ and\ \citenamefont {Heyl}}]{Halimeh2021}%
  \BibitemOpen
  \bibfield  {author} {\bibinfo {author} {\bibfnamefont {J.~C.}\ \bibnamefont {Halimeh}}, \bibinfo {author} {\bibfnamefont {D.}~\bibnamefont {Trapin}}, \bibinfo {author} {\bibfnamefont {M.}~\bibnamefont {Van~Damme}},\ and\ \bibinfo {author} {\bibfnamefont {M.}~\bibnamefont {Heyl}},\ }\bibfield  {title} {\bibinfo {title} {Local measures of dynamical quantum phase transitions},\ }\href {https://doi.org/10.1103/PhysRevB.104.075130} {\bibfield  {journal} {\bibinfo  {journal} {Physical Review B}\ }\textbf {\bibinfo {volume} {104}},\ \bibinfo {pages} {075130} (\bibinfo {year} {2021})}\BibitemShut {NoStop}%
\bibitem [{\citenamefont {Troyer}\ and\ \citenamefont {Wiese}(2005)}]{Troyer2005}%
  \BibitemOpen
  \bibfield  {author} {\bibinfo {author} {\bibfnamefont {M.}~\bibnamefont {Troyer}}\ and\ \bibinfo {author} {\bibfnamefont {U.-J.}\ \bibnamefont {Wiese}},\ }\bibfield  {title} {\bibinfo {title} {Computational complexity and fundamental limitations to fermionic quantum {Monte Carlo} simulations},\ }\href {https://doi.org/10.1103/PhysRevLett.94.170201} {\bibfield  {journal} {\bibinfo  {journal} {Physical Review Letters}\ }\textbf {\bibinfo {volume} {94}},\ \bibinfo {pages} {170201} (\bibinfo {year} {2005})}\BibitemShut {NoStop}%
\bibitem [{\citenamefont {Huang}\ \emph {et~al.}(2020)\citenamefont {Huang}, \citenamefont {Kueng},\ and\ \citenamefont {Preskill}}]{Huang2020}%
  \BibitemOpen
  \bibfield  {author} {\bibinfo {author} {\bibfnamefont {H.-Y.}\ \bibnamefont {Huang}}, \bibinfo {author} {\bibfnamefont {R.}~\bibnamefont {Kueng}},\ and\ \bibinfo {author} {\bibfnamefont {J.}~\bibnamefont {Preskill}},\ }\bibfield  {title} {\bibinfo {title} {Predicting many properties of a quantum system from very few measurements},\ }\href {https://doi.org/10.1038/s41567-020-0932-7} {\bibfield  {journal} {\bibinfo  {journal} {Nature Physics}\ }\textbf {\bibinfo {volume} {16}},\ \bibinfo {pages} {1050} (\bibinfo {year} {2020})}\BibitemShut {NoStop}%
\end{thebibliography}

\begin{thebibliography}{3}%
\makeatletter
\providecommand \@ifxundefined [1]{%
 \@ifx{#1\undefined}
}%
\providecommand \@ifnum [1]{%
 \ifnum #1\expandafter \@firstoftwo
 \else \expandafter \@secondoftwo
 \fi
}%
\providecommand \@ifx [1]{%
 \ifx #1\expandafter \@firstoftwo
 \else \expandafter \@secondoftwo
 \fi
}%
\providecommand \natexlab [1]{#1}%
\providecommand \enquote  [1]{``#1''}%
\providecommand \bibnamefont  [1]{#1}%
\providecommand \bibfnamefont [1]{#1}%
\providecommand \citenamefont [1]{#1}%
\providecommand \href@noop [0]{\@secondoftwo}%
\providecommand \href [0]{\begingroup \@sanitize@url \@href}%
\providecommand \@href[1]{\@@startlink{#1}\@@href}%
\providecommand \@@href[1]{\endgroup#1\@@endlink}%
\providecommand \@sanitize@url [0]{\catcode `\\12\catcode `\$12\catcode `\&12\catcode `\#12\catcode `\^12\catcode `\_12\catcode `\%12\relax}%
\providecommand \@@startlink[1]{}%
\providecommand \@@endlink[0]{}%
\providecommand \url  [0]{\begingroup\@sanitize@url \@url }%
\providecommand \@url [1]{\endgroup\@href {#1}{\urlprefix }}%
\providecommand \urlprefix  [0]{URL }%
\providecommand \Eprint [0]{\href }%
\providecommand \doibase [0]{https://doi.org/}%
\providecommand \selectlanguage [0]{\@gobble}%
\providecommand \bibinfo  [0]{\@secondoftwo}%
\providecommand \bibfield  [0]{\@secondoftwo}%
\providecommand \translation [1]{[#1]}%
\providecommand \BibitemOpen [0]{}%
\providecommand \bibitemStop [0]{}%
\providecommand \bibitemNoStop [0]{.\EOS\space}%
\providecommand \EOS [0]{\spacefactor3000\relax}%
\providecommand \BibitemShut  [1]{\csname bibitem#1\endcsname}%
\let\auto@bib@innerbib\@empty
\bibitem [{\citenamefont {Gross}(2006)}]{supp:Gross2006}%
  \BibitemOpen
  \bibfield  {author} {\bibinfo {author} {\bibfnamefont {D.}~\bibnamefont {Gross}},\ }\bibfield  {title} {\bibinfo {title} {Hudson's theorem for finite-dimensional quantum systems},\ }\href {https://doi.org/10.1063/1.2393152} {\bibfield  {journal} {\bibinfo  {journal} {Journal of Mathematical Physics}\ }\textbf {\bibinfo {volume} {47}},\ \bibinfo {pages} {122107} (\bibinfo {year} {2006})}\BibitemShut {NoStop}%
\bibitem [{\citenamefont {Vidal}(2004)}]{supp:Vidal2004}%
  \BibitemOpen
  \bibfield  {author} {\bibinfo {author} {\bibfnamefont {G.}~\bibnamefont {Vidal}},\ }\bibfield  {title} {\bibinfo {title} {Efficient simulation of one-dimensional quantum many-body systems},\ }\href {https://doi.org/10.1103/PhysRevLett.93.040502} {\bibfield  {journal} {\bibinfo  {journal} {Physical Review Letters}\ }\textbf {\bibinfo {volume} {93}},\ \bibinfo {pages} {040502} (\bibinfo {year} {2004})}\BibitemShut {NoStop}%
\bibitem [{\citenamefont {Vidal}(2007)}]{supp:Vidal2007Infinite}%
  \BibitemOpen
  \bibfield  {author} {\bibinfo {author} {\bibfnamefont {G.}~\bibnamefont {Vidal}},\ }\bibfield  {title} {\bibinfo {title} {Classical simulation of infinite-size quantum lattice systems in one spatial dimension},\ }\href {https://doi.org/10.1103/PhysRevLett.98.070201} {\bibfield  {journal} {\bibinfo  {journal} {Phys. Rev. Lett.}\ }\textbf {\bibinfo {volume} {98}},\ \bibinfo {pages} {070201} (\bibinfo {year} {2007})},\ \Eprint {https://arxiv.org/abs/cond-mat/0605597} {arXiv:cond-mat/0605597} \BibitemShut {NoStop}%
\end{thebibliography}
\end{document}